\documentclass{statsoc}

\usepackage[a4paper]{geometry}
\usepackage{graphicx}
\usepackage[textwidth=8em,textsize=small]{todonotes}
\usepackage{amsmath}
\usepackage{natbib}
\usepackage{amssymb}
\usepackage{amsfonts}
\usepackage{enumerate}
\usepackage{booktabs}
\usepackage{multirow}
\usepackage{hyperref}
\usepackage{color}
\usepackage{bm}
\usepackage{bbm}
\usepackage{parskip}
\usepackage{tikz}
\usepackage[export]{adjustbox}
\usepackage{url} 
\usepackage{subcaption}
\usepackage{mathtools}
\usepackage{floatrow}
\usepackage[toc,page]{appendix}
\let\proof\relax
\let\endproof\relax
\usepackage{amsthm}

\title[Veracity scoring in geostatistics]{On Statistical Properties of A Veracity Scoring Method for Spatial Data}
\author[A. Chakraborty, S. N. Lahiri]{Arnab Chakraborty$^1$, Soumendra N. Lahiri$^1$}
\address{$^{1}$ North Carolina State University (Department of Staistics)}

\begin{document}

\newcommand{\rn}{\mbox{${\mathbb{R}^{n}}$}}
\newcommand{\rtwo}{\mbox{${\mathbb{R}^{2}}$}}
\newcommand{\itwo}{\mbox{${\mathbb{I}^{2}}$}}
\newcommand{\rone}{\mbox{${\mathbb{R}}$}}
\newcommand{\rplus}{\mbox{${\mathbb{R}^{n+1}}$}}
\newcommand{\rrm}{\mbox{${\mathbb{R}^{m}}$}}
\newcommand{\rp}{\mbox{${\mathbb{R}^{p}}$}}
\newcommand{\rd}{\mbox{${\mathbb{R}^{d}}$}}
\newcommand{\rdXm}{\mbox{${\mathbb{R}^{d\times m}}$}}
\newcommand{\rk}{\mbox{${\mathbb{R}^{k}}$}}
\newcommand{\var}{\mathrm{Var}}
\newcommand{\sgn}{\mathrm{sgn}}
\newcommand{\likesingle}{\mbox{$\frac{f_1(x_i)}{f_0(x_i)} $}}
\newcommand{\rpm}{\sbox0{$1$}\sbox2{$\scriptstyle\pm$}
  \raise\dimexpr(\ht0-\ht2)/2\relax\box2 }
\newcommand{\bfs}{\mbox{${\mathbf{s}}$}}
\newcommand{\bft}{\mbox{${\mathbf{t}}$}}
\newcommand{\bfh}{\mbox{${\mathbf{h}}$}}
\newcommand{\bfx}{\mbox{${\mathbf{x}}$}}
\newcommand{\bfe}{\mbox{${\mathbf{e}}$}}
\newcommand{\bfg}{\mbox{${\mathbf{g}}$}}
\newcommand{\bfw}{\mbox{${\mathbf{w}}$}}
\newcommand{\bfu}{\mbox{${\mathbf{u}}$}}
\newcommand{\bfX}{\mbox{${\mathbf{X}}$}}
\newcommand{\bfW}{\mbox{${\mathbf{W}}$}}
\newcommand{\bfA}{\mbox{${\mathbf{A}}$}}
\newcommand{\bfU}{\mbox{${\mathbf{U}}$}}
\newcommand{\bfI}{\mbox{${\mathbf{I}}$}}
\newcommand{\bfH}{\mbox{${\mathbf{H}}$}}
\newcommand{\bfC}{\mbox{${\mathbf{C}}$}}
\newcommand{\bfF}{\mbox{${\mathbf{F}}$}}
\newcommand{\bfbeta}{\mbox{$\boldsymbol{\beta}$}}
\newcommand{\bfeta}{\mbox{$\boldsymbol{\eta}$}}
\newcommand{\bfmu}{\mbox{$\boldsymbol{\mu}$}}
\newcommand{\bfnu}{\mbox{$\boldsymbol{\nu}$}}
\newcommand{\bfeps}{\mbox{$\boldsymbol{\epsilon}$}}
\newcommand{\bfgamma}{\mbox{$\boldsymbol{\gamma}$}}
\newcommand{\bfGamma}{\mbox{$\boldsymbol{\Gamma}$}}
\newcommand{\bftheta}{\mbox{$\boldsymbol{\theta}$}}
\newcommand{\bfalpha}{\mbox{$\boldsymbol{\alpha}$}}
\newcommand{\bfomega}{\mbox{$\boldsymbol{\omega}$}}
\newcommand{\atant}{\mathrm{atan2}}
\newcommand{\imi}{\mathrm{i}}
\renewcommand{\qedsymbol}{$\blacksquare$}
\def\Lp{\left(}
\def\Rp{\right)}
\def\LP{\left\{ } 
\def\RP{\right\}}

\theoremstyle{plain}
\newtheorem{theo}{Theorem}
\newtheorem{cor}{Corollary}
\newtheorem{lem}{Lemma}
\newtheorem{prop}{Proposition}
\newtheorem{def1}{Definition}[section]
\newtheorem{claim}{Claim}

\theoremstyle{definition}
\newtheorem{eg}{Example}

\theoremstyle{definition}
\newtheorem{rem}{Remark}

\numberwithin{equation}{section}
\numberwithin{theo}{section}
\numberwithin{cor}{section}
\numberwithin{lem}{section}
\numberwithin{prop}{section}
\numberwithin{cor}{section}
\numberwithin{def1}{section}
\numberwithin{rem}{section}
\numberwithin{equation}{section}


\def\Lp{\left(}
\def\Rp{\right)}
\def\LP{\left\{ } 
\def\RP{\right\}}

\begin{abstract}
Measuring veracity or reliability of noisy data is of utmost importance, especially in the scenarios where the information are gathered through automated systems. In a recent paper, \cite{chak} have introduced a veracity scoring technique for geostatistical data. The authors have used a high-quality `reference' data to measure the veracity of the varying-quality observations and incorporated the veracity scores in their analysis of mobile-sensor generated noisy weather data to generate efficient predictions of the ambient temperature process. In this paper, we consider the scenario when no reference data is available and hence, the veracity scores (referred as VS) are defined based on `local' summaries of the observations. We develop a VS-based estimation method for parameters of a spatial regression model. Under a non-stationary noise structure and fairly general assumptions on the underlying spatial process, we show that the VS-based estimators of the regression parameters are consistent. Moreover, we establish the advantage of the VS-based estimators as compared to the ordinary least squares (OLS) estimator by analyzing their asymptotic mean squared errors. We illustrate the merits of the VS-based technique through simulations and apply the methodology to a real data set on \textit{mass percentages} of ash in coal seams in Pennsylvania.

\end{abstract}

\section{Introduction}
\label{sec:intro}
With the advancement of sensor-related technologies in the field of data science and big-data analytics, `veracity' is becoming one of the most important V's, along with velocity, variety and volume. In a recent article, \cite{luko14} mentioned how the data source, data collection technologies and data processing methodologies could induce bias, ambiguity and noise in real-world big-data applications. Practitioners often not only want to detect the noisy observations but also want to assign a reliability score to each of the observations, which efficiently indicates the `relative' magnitude of the associated noise with it. For example, studies conducted by \cite{evans97} show that dynamic display of veracity of the contents can help the map-users (e.g. in google maps) to take advantage of the reliability information in making decisions. Though there are some works on veracity analysis in media and communications (for example, see \citealt{conroy15}; \citealt{rendon18} etc.) reliability assessment in the analysis of spatial data is not common in literature. Standard geostatistical analysis of the corrupted observations without taking the noisy nature of the data into account can result in erroneous inference and prediction. In a recent study, \cite{chak} introduced a reliability metric, namely veracity score (VS), to assess the credibility of the varying-quality crowdsourced observations in a geostatistical setting. In their analysis of mobile-sensor-generated noisy weather data, the authors have used a `high-quality' reference data -- coming from the ground weather stations -- to define the veracity scores (VS) of the noisy observations. 
However, often in practice, such high-quality reference data is not available. In this paper, we consider the VS when there is no reference data available. 

Although \cite{chak} have proposed veracity scoring based on `local' summaries from the noisy data, statistical properties of the VS and corresponding VS-based estimators remain unexplored. In this paper, we investigate statistical properties of the VS-based methodology under fairly general conditions on the underlying spatial process. We assume an additive-multiplicative noise structure for the corrupted observations. In contrast to the more commonly used i.i.d. additive measurement errors (see e.g., \citealt{diggle10}), the underlying noise structure here is non-stationary due to the presence of the multiplicative component. Assuming a suitable spatial asymptotic framework (e.g. mixed-increasing domain, \citealt{hall94}), we establish the consistency of the VS-based regression parameter estimator. The key result used here is a simplified asymptotic representation of the VS based on an extension of the Ghosh-Bahadur representation of sample quantiles of irregularly spaced spatial data. In addition, we show that the MSE of the VS-based estimator does not depend on the noise variances associated with the `bad' observations asymptotically. This explains the empirical observation on robustness of the VS-based estimator against the noise inherent in crowdsourced spatial data as reported in \cite{chak}. We also considered asymptotic behavior of the MSE of the OLS estimator. We show that the MSE of the OLS estimator is bounded below by a quantity that involves the sum of the associated noise variances. In situations where the combined noise level of the `bad' observations is high, accuracy of the OLS estimator can be substantially inferior to the VS-based estimator.

Next, we evaluate finite sample properties of VS-based estimation though extensive simulations. As the proportion of noisy observations increases from $5\%$ to $20\%$, the relative efficiency (defined as ratio of MSEs) of the VS-based estimators of the regression parameters increases from $145\%$ to more than $300\%$ for a sample of size $500$ (see Table~\ref{tab:RegParamMSE}). Similar advantages of the VS-based estimators are noted at varying levels of the additive and multiplicative noise variances. We also carried out a comparative numerical analysis with the robust-REML method, a competing robust geostatistical technique proposed by \cite{kunsch11} (also, see \citealt{georob18}, \citealt{georobMan}). The VS-based technique outperforms the robust-REML method both in terms of scalability and statistical accuracy (see Table~\ref{tab:VSvsGR}). Finally, we applied the VS-based analysis on a real data set containing measurements of mass percentages of ash in coal seams. The VS-based methodology identified all of the outliers previously detected in \cite{cressie93} and \cite{georobMan}'s analysis of the coal ash data. More importantly, our analysis indicates the presence of spatial correlation (see Table~\ref{tab:coalashCovEst}) while the earlier analyses completely failed to capture the spatial dependence. Incorporation of the spatial dependence in the VS-based analysis led to comparable or much smaller point-wise prediction errors (indicated by the sizes of the circles in Figure~\ref{fig:re-vs-gr}b).

The organization of the rest of the paper is as follows. In Section~\ref{sec:vs}, we provide the background, introduce some notation, and briefly review the VS-based methodology. Section~\ref{sec:AsympVS} defines the theoretical framework, states the assumptions and the results regarding asymptotic properties of the VS-based as well as the OLS estimators. In Section~\ref{sec:SimStudy}, we summarize the simulation design and simulation results for evaluating the VS-based methodology. Section~\ref{sec:readl-dat-ex} provides the details of the real data example. Finally, in Section~\ref{sec:conclusion}, we summarize our findings and future directions to this work. Proofs of the main results of this paper are given in Appendix~\ref{sec:proofs}. A supplementary material of this paper is available which contains the proofs of the auxiliary results and additional simulation results.

\section{Veracity Score (VS) Methods}
\label{sec:vs}
In this section, we briefly review the VS-based method in geostatistics. Before that, here we provide the background and a brief recap of some of the preliminary notation introduced in \cite{chak}.

Let $\LP Y(\bfs): \bfs \in \mathcal{D} \subset \rtwo \RP$ be the spatial process of our interest. We assume a spatial regression model, i.e. $\LP Y(\bfs) \RP$ has a decomposition of the form
\begin{equation}
\label{eq:LinMod}
    \begin{split}
        Y(\bfs) = \bfx(\bfs)^\prime \bfbeta + \epsilon(\bfs),
    \end{split}
\end{equation} where $\bfx(\cdot) = \Lp x_1(\cdot),...,x_p(\cdot) \Rp^\prime$ is the $p$-dimensional deterministic vector process of known covariates, $\bfbeta$ is the unknown regression parameter vector, and $\LP \epsilon(\bfs) \RP$ is a spatially correlated residual process. We assume that $\LP \epsilon(\cdot) \RP$ is an intrinsically stationary process with an admissible parametric variogram function $
2\gamma(\bfh; \bftheta) = \text{Var}\LP \epsilon(\bfs) - \epsilon(\bfs + \bfh) \RP
$ where $\bftheta$ is the covariance parameter of interest. Under second-order stationarity, it can be shown that $\gamma(\bfh) = C(\boldsymbol{0}) - C(\bfh) $ where $C(\bfh) = \text{Cov}\Lp \epsilon(\bfs), \epsilon(\bfs + \bfh) \Rp$ is the covariance function of the $\epsilon$-process. For more details about variograms and covariograms see Chapter 2 in \cite{cressie93}.

Now, instead of observing $\LP Y(\bfs_1), \dots , Y(\bfs_n) \RP$ at locations (can be irregularly space) $\mathcal{S}_n = \LP \bfs_1, \dots , \bfs_n \RP$, we observe corrupted observations $\LP Z(\bfs_1), \dots , Z(\bfs_n) \RP$. We write the corrupted observations as,
\begin{equation} \label{eq:genMod_Z}
\begin{split}
Z(\bfs_i) = \bfx(\bfs_i)^\prime \bfbeta + w(\bfs_i),
\end{split}
\end{equation} where $\LP w(\bfs) \RP$ is a mean-zero spatially correlated process, possibly nonstationary, which contains noise in addition to the small-scale local variations induced by $\epsilon(\bfs)$. For example, as mentioned by \cite{chak}, let the corruption is coming through an additive-multiplicative noise model, i.e.
\begin{equation} \label{eq:AddMultNoiseMod}
\begin{split}
Z(\bfs_i) = \epsilon_{M_i} Y(\bfs_i) + \epsilon_{A_i}(\bfs_i),
\end{split}
\end{equation} where $\epsilon_{M_i}$ and $\epsilon_{A_i}$ are the randomm variables corresponding to the multiplicative and additive noise parts. Then the $w$-process can be written as $w(\bfs_i) = \epsilon_{M_i}(\mu(\bfs_i) - 1) + \epsilon_{M_i}\epsilon(\bfs_i) + \epsilon_{A_i}$. In case there is no multiplicative component, $w(\bfs_i) = \epsilon(\bfs_i) + \epsilon_{A_i}$. Observe that existance of the multiplicative component make the noisy residuals $\LP w(\bfs) \RP$ dependent on the location $\bfs$ and hence non-stationary in nature.

\subsection{Formulation of VS}
\label{subsec:Formulation-VS}
Denote a square $\delta$-neighborhood ($\delta > 0$) around location $\bfs$ as $\mathcal{B}_{\delta}(\bfs)$, i.e. $\mathcal{B}_{\delta}(\bfs) = (\bfs - \delta, \bfs + \delta]$, where the subtractions are component-wise. Then the VS of the observation $Z(\bfs_i)$ is defined as \citep{chak},
\begin{equation}\label{eq:VerSc_woRef}
\begin{split}
V(\bfs_i) = \phi\Lp \frac{\lvert Z(\bfs_i) - \mathcal{C}(\mathbf{Z}_i) \rvert}{\alpha + D(\mathbf{Z}_i)}\Rp,
\end{split}
\end{equation}
where $\phi : \rone^+\cup \LP 0 \RP \to \rone^+ \cup \LP 0 \RP$ is some non-increasing bounded above function, referred as \textit{veracity function}; $\alpha \geq 0$ is a regularity parameter, called the \textit{baseline deviation}; $\mathbf{Z}_i = \Lp Z(\bfs_{i_1}), \dots , Z(\bfs_{i_{n(i)}})\Rp^\prime$ where $\{ \bfs_{i_1}, \dots , \bfs_{i_{n(i)}} \}$ is the set of observation locations in the small $\delta$-neighborhood $\mathcal{B}_\delta(\bfs_i)$; and finally, $\mathcal{C}(\mathbf{x})$ is a measure of central tendency and $D(\mathbf{x})$ is a measure of dispersion of the values in the vector $\mathbf{x}$. Clearly, the definition of VS given in Equation~\ref{eq:VerSc_woRef} measures the amount of deviation of the observation $Z(\bfs_i)$ from the `local' summary $\mathcal{C}(\mathbf{Z}_i)$ relative to the `local' variation $D(\mathbf{Z}_i)$. The veracity function $\phi(\cdot)$ being a non-increasing function, larger the value of the scaled deviation $\frac{\lvert Z(\bfs_i) - \mathcal{C}(\mathbf{Z}_i) \rvert}{\alpha + D(\mathbf{Z}_i)}$ lower is the VS and for lower values of the scaled deviation, VS will be higher.

For the measure of center and dispersion we have considered the following two choices that are widely used by practitioners:
\begin{enumerate}
    \item $\mathcal{C}(\mathbf{Z}_i) = Q_2(\mathbf{Z}_i)$ and $D(\mathbf{Z}_i) = \text{IQR}(\mathbf{Z}_i) = Q_3(\mathbf{Z}_i) - Q_1(\mathbf{Z}_i)$, where $Q_j(\bfx)$ denotes the $j$-th quartile of the observations in $\bfx$, $j \in \LP 1, 2, 3, 4 \RP$;
    \item $\mathcal{C}(\mathbf{Z}_i) = \bar{Z}_{i_.} = \frac{1}{n(i)}\sum_{j = 1}^{n(i)} Z(\bfs_{i_j})$, and $D(\mathbf{Z}_i) = \text{sd}(\mathbf{Z}_i) = \sqrt{\frac{1}{n(i)-1}\sum_{j} \Lp Z(\bfs_{i_j}) - \bar{Z}_{i_.} \Rp^2}$.
\end{enumerate} Clearly, the first set of choices are based on sample quantiles and are expected to be more robust and efficient as compared to the sample mean and standard deviation in the analysis of noisy data (for details, see \citealt{sen68}). In this paper, we refer the VS with choice (a) as the measure of location and scale as `Medain-VS' and denote it by $V^{(m)}(\bfs_i)$. Similarly, for choice (b) we denote the VS as $V^{(a)}(\bfs_i)$ ($(a)$ for average) and refer it as `Mean-VS'. For the veracity function we have adopted the choice used by \cite{chak}, i.e. $\phi(x) = \exp(-x)$.

\subsection{VS-based estimation in spatial regression}
\label{subsec:VS-based-est}
Under the setting discussed above, as the covariance parameter is unknown, the standard practice is to estimate the regression parameter using \textit{ordinary least squares} (OLS): $
\hat{\bfbeta}_{\text{ols}} = \Lp \mathbf{X}^\prime \mathbf{X} \Rp^{-1} \mathbf{X}^\prime\mathbf{Z},
$ where $\mathbf{X} = \Lp \bfx(\bfs_1), \dots , \bfx(\bfs_n) \Rp^\prime$. Next, the residuals from the OLS fit are used in \textit{weighted least squares}-based variogram estimation (for details, see page 90, Chapter 2, \citealt{cressie93}) to obtain the covariance parameter estimator $\hat{\bftheta}_{\text{wls}}$. Once the covariance parameter is estimated, one can try to improve the mean parameter estimates by using \textit{estimated generalized least squares} (EGLS) estimator, given by $\hat{\bfbeta}_{\text{egls}} = \Lp X^\prime \Sigma^{-1}(\hat{\bftheta}_{\text{wls}}) X \Rp^{-1} X^\prime \Sigma^{-1}(\hat{\bftheta}_{\text{wls}}) \mathbf{Z}$, where $\Sigma(\hat{\bftheta}_{\text{wls}})$ is the estimated covariance matrix. But, this last stage of updating the OLS fit introduces additional variability due to the estimation of the covariance parameters and hence not necessarily preferable than the OLS estimator. Finally, the estimated mean and covariance structure is used to predict the process at a new location $\bfs_0$ using \textit{best linear unbiased predictor} or kriging predictor (for details, see Chapter 3, \citealt{cressie93}).

The standard approach of estimation and kriging is not robust in nature and hence, may produce erroneous inference and prediction in case there are outliers in the data. To estimate the mean parameter $\bfbeta$ robustly, instead of using ordinary squared error loss, in VS-based estimation, weighted squared error loss is used with VS as the corresponding weights \citep{chak} as shown below:
\begin{equation}\label{eq:VSEstimator}
\begin{split}
\hat{\bfbeta}_{\text{vs}} &= \underset{\bfbeta}{\text{argmin}} \; \sum_{i = 1}^n V(\bfs_i)\Lp  Z(\bfs_i)-  \mathbf{x}(\bfs_i)^\prime \bfbeta \Rp^2\\
&= \Lp \bfX^\prime \mathbf{D}_v \bfX \Rp^{-1}\bfX^\prime \mathbf{D}_v \mathbf{Z},
\end{split}
\end{equation} where $\mathbf{D}_v = \text{diag}( V(\bfs_1), \dots $ $, V(\bfs_n) )$. Clearly, as the VS is expected to be smaller for `bad' observations, the VS-based estimator given in Equation~\ref{eq:VSEstimator} is expected to be affected less by the high noise associated with those. If the VS can capture the noisy observations perfectly, i.e. $V(\bfs_i) \approx 0$ if $Z(\bfs_i)$ is contaminated and $V(\bfs_i) = 1$ otherwise, then $\hat{\bfbeta}_{\text{vs}}$ is approximately equal to the OLS estimator computed only from the `noiseless' observations in the data.

Depending on the versions of VS used to compute the VS-based estimator, we consider two versions of the VS-based estimator in this article: `Mean-VS', denoted by $\hat{\bfbeta}^{(a)}_{\text{vs}}$, where the sample mean and standard deviations are used to assess veracity and `Median-VS', denoted by $\hat{\bfbeta}^{(m)}_{\text{vs}}$, with sample median and IQR in the definition of VS. Though the focus of this work is mainly on the properties of the `Median-VS' regression estimator, in the following sections we put some remarks about the properties and asymptotic behaviors of the `Mean-VS' version as well.

The de-trended observations, i.e. $\hat{\epsilon}_{\text{vs}}(\bfs_i) = Z(\bfs_i) - \mathbf{x}(\bfs_i)^\prime \hat{\bfbeta}_{\text{vs}}$ for $i \in \LP1,2,...,n\RP$ -- unlike the case of dealing with high-quality observations -- are not free of noise and hence, it is not reasonable to use these directly in the covariance parameter estimation. \cite{chak} proposed a VS-based smoothing of the residuals:
\begin{equation}\label{eq:NewRes_wRef}
\begin{split}
\tilde{\epsilon}(\bfs_i) = V(\bfs_i)^q \hat{\epsilon}_{\text{vs}}(\bfs_i) + (1 - V(\bfs_i)^q)Q_2(\mathbf{Z}_i - \mathbf{X}_i \hat{\bfbeta}_{\text{vs}}),
\end{split}
\end{equation} where $\mathbf{X}_i = \Lp \bfx(\bfs_{i_1}),...,\bfx(\bfs_{i_{n(i)}}) \Rp^\prime$ is the $n(i)\times p$ matrix of the covariates corresponding to the observations in $\mathcal{B}_{\boldsymbol{\delta}}(\bfs_i)$; and $q \geq 0$ is another regularity parameter which determines the smoothness of the residuals defined in Equation~\ref{eq:NewRes_wRef}. Next, these VS-based smoothed residual vector $\tilde{\bfeps} = \Lp \tilde{\epsilon}(\bfs_1), \dots , \tilde{\epsilon}(\bfs_n) \Rp^\prime$ is used in least squares based variogram model fitting (for example, see \citealt{cressie80}) to get the VS-based covariance parameter estimator $\hat{\bftheta}_{\text{vs}}$. Finally, the prediction of the $\epsilon$-process at a new location $\bfs_0$, denoted by $\tilde{\epsilon}(\bfs_0)$, is obtained using ordinary kriging with the estimated covariance parameters and the VS-based smoothed residuals as discussed in \cite{chak}.

In the next section, we have focused on the asymptotic behavior of the VS and VS-based regression parameter estimator. Though investigation of asymptotic properties VS-based covariance estimation is beyond the scope of this paper, we have evaluated the same using simulations in Section~\ref{sec:SimStudy}.

\section{Asymptotic properties of the VS-based regression estimator}
\label{sec:AsympVS}
Before going to the main results of this section we need to introduce some notation, specify the theoretical framework and state the regularity conditions. From this section onward, by $C(\cdot), C, C_1, C_2,\dots$ we denote constants with respect to the sample size $n$. For $d \geq 2$, we denote the volume of a set $A \subset \rd$ as $\lvert A \rvert$, i.e., the Lebesgue measure of $A$ if it has nonzero volume and the cardinality of $A$ if $A$ is finite. For two sequences of positive reals $\LP a_n \RP_{n = 1}^\infty$ and $\LP b_n \RP_{n = 1}^\infty$ we say one is \textit{of the order} of another (denoted as $a_n \sim b_n$) if $a_n/b_n \to C > 0$.

\subsection{Model specification}
\label{subsec:ModelSpec}
As discussed before, the process of our interest is denoted as $\LP Y(\bfs) : \bfs \in \mathcal{D} \subset \rtwo \RP$ and instead of observing realizations from the `true' process, we observe corrupted observations $\LP Z(\bfs_1), \dots , Z(\bfs_n) \RP$ where, the observations locations $\LP \bfs_1, \dots , \bfs_n \RP$ are possibly irregularly spaced. The results of this paper are specific to the spatial regression model defined in Equation~\ref{eq:LinMod}. For the zero mean second-order stationary residual process $\LP \epsilon(\bfs) : \bfs \in \mathcal{D}\RP$, let us denote the marginal distribution function of $\epsilon(\bfs)$ as $F_\epsilon(\frac{x}{\sigma_\epsilon})$ and, the spatial dependence structure can be formulated as,
$$
\text{Cov}\Lp \epsilon(\bfs_i), \epsilon(\bfs_j) \Rp = \begin{cases} \sigma_\epsilon^2 \; \rho_{\epsilon} \Lp \bfs_i - \bfs_j \Rp \;\; &\text{if} \;\; \bfs_i \neq \bfs_j\\
\sigma_\epsilon^2 \;\; &\text{otherwise.}  \end{cases}
$$ Here, $F_\epsilon(\cdot)$ is a distribution function of a zero mean unit-variance random variable and $\rho_\epsilon(\cdot)$ is a non-negative definite function on $\rtwo$ with $\rho_\epsilon(\mathbf{0}) = 1$. To preserve the identifiability, we assume that the  supports of $\LP \epsilon_{M_i}: i \in  \LP 1, \dots , n \RP \RP$ are contained in $[0,\infty)$.

We further assume that among $\LP Z(\bfs_1), \dots , Z(\bfs_n) \RP$ only a `small' portion is `bad-quality data' and the corrupted observations are coming through the additive-multiplicative noise model given in Equation~\ref{eq:AddMultNoiseMod}. We formulate this assumption as follows.
\begin{equation}
\label{eq:error_dist}
\begin{split}
\epsilon_{M_i} \sim \begin{cases}
 \Delta(1) \;\; &\text{if}\;\; i \in G_n\\
 F_M\Lp \frac{\cdot - 1}{\sigma_M} \Rp \;\; &\text{o.w.}
 \end{cases}\; ;\;\;
\epsilon_{A_i}  \sim \begin{cases}
 \Delta(0) \;\; &\text{if}\;\; i \in G_n\\
 F_A\Lp \frac{\cdot}{\sigma_A} \Rp \;\; &\text{o.w.}\; ,
\end{cases}
\end{split}
\end{equation} where, $\Delta(x)$ denotes a degenerate distribution with point mass at $-\infty < x < \infty$; $G_n \subset \LP 1, \dots , n \RP$ is a subset of indices; and $\sigma_M, \sigma_A$ are positive constants. With this model, if $i \in G_n$, we have no noise associated with the observation, i.e., $Z(\bfs_i) = Y(\bfs_i)$. If $i \notin G_n$, then $Z(\bfs_i) = \epsilon_{M_i} Y(\bfs_i) + \epsilon_{A_i}$, where $\epsilon_{M_i}$ and $\epsilon_{A_i}$ have positive variances. Also, we have taken $\LP \epsilon_{M_i}\RP_{i = 1}^n$, $\LP \epsilon_{A_i}\RP_{i = 1}^n$ and $\LP \epsilon(\bfs_i) \RP_{i = 1}^n$ are independent of each other. We further assume that the proportion of ``good'' observations is a constant (w.r.t. $n$) denoted by $q_e$, i.e., $\lvert G_n \rvert /n \approx q_e$, and $1 - q_e$ is the proportion of noisy observations in the data. Observe that, the variance corresponding to $\epsilon_{M_i}$, $\sigma_{M_i} = \sigma_M$ if $i \notin G_n$ and $\sigma_{M_i} = 0$ if $i \in G_n$. Similar observation can be made for the additive noise variances (denoted by $\sigma_{A_i}$) as well.

Under the specified model we can rewrite our observations as,
\begin{equation}
    \begin{split}
        Z(\bfs_i) = \bfx(\bfs_i)^\prime \bfbeta + w(\bfs_i),
    \end{split}
\end{equation} where $E(w(\bfs_i)) = 0$, $\var(w(\bfs_i)) = \sigma_i^2$ and,
$$
\begin{aligned}
\sigma_i^2  &= \begin{cases}
 \sigma_\epsilon^2 \;\; &\text{if}\;\; i \in G_n\\
  \sigma_\epsilon^2 + \tau_i^2\;\; &\text{o.w.},
 \end{cases}
\end{aligned}
$$ where, $\tau_i^2 \equiv \tau(\bfs_i)^2 = (\bfx(\bfs_i)^\prime \bfbeta)^2 \sigma_M^2 + \sigma_\epsilon^2\sigma_M^2 + \sigma_A^2$ is the additional noise variance associated with the corrupted observation at location $\bfs_i$. Clearly, if $i \in G_n$, $\tau_i^2 = 0$. We denote the marginal distribution function of $w(\bfs_i)$ as $F_i(x)$ for $i \in \LP 1, \dots , n\RP$, for any $n \geq 2$. Clearly, if $i \in G_n$, $w(\bfs_i) =\epsilon(\bfs_i)$ and $F_i(x) = F_\epsilon(x/\sigma_\epsilon)$ and if $i \in G_n^c$, we denote the marginal distribution function of $w(\bfs_i)$ as $F_2(\frac{x}{\sigma_i})$, where $F_2(\cdot)$ is again a distribution function of a centered and scaled random variable. Without loss of any generality, we take the baseline deviation parameter $\alpha$ in the definition of VS to be equal to $0$. The results of this paper can be straightforwardly extended for any other finite constant (w.r.t. $n$) $\alpha > 0$.

\subsection{Spatial framework and notations}
We take the sampling region to be $\mathcal{D} \equiv \mathcal{D}_n = \lambda_n [0,1]^2$, i.e. a 2-dimensional square region with area $\lambda_n^2$ where $\LP \lambda_n \RP$ is a sequence of positive reals such that $\lambda_n \to \infty$ as $n \to \infty$ and 
\begin{equation}
\label{eq:CondnLambda}
\begin{split}
\lambda_n^{-1}  + n^{-1} \lambda_n^{2} \to 0.
\end{split}
\end{equation} Under condition (\ref{eq:CondnLambda}), the spatial asymptotic framework we consider is similar to the \textit{mixed increasing-domain} asymptotic framework used in \cite{hall94}, \cite{lahiri99}, etc.The first component of (\ref{eq:CondnLambda}) states the domain has to increase with the sample size, and the second component allows the possibility of infilling sampling points as well. The results stated here are not particular to the shape and position of the rectangular region $\mathcal{D}_n$. Our results can be extended to more general shapes using the asymptotic framework where the region is obtained by inflating a prototype region contained in $(-1/2, 1/2]^2$ with center at the origin by a scaling constant as discussed in \cite{lahiri02}.

Now, we introduce some notation required to state the regularity conditions, the results as well as the proofs of this paper. For any $A \subset \rtwo$ and a random field $\LP T(\bfs): \bfs \in \rtwo \RP$ let us denote $\mathcal{F}_T(A) = \sigma \langle T(\bfs) : \bfs \in A \rangle$, the $\sigma$-field generated by the random variables $\LP T(\bfs) : \bfs \in A \RP$. For any $d \geq 2$ and for any $\mathbf{x} \in \rd$ the $L_1$ and $L_2$ norms are denoted by $\lvert \mathbf{x} \rvert$ and $\lVert \mathbf{x} \rVert$ respectively. For any two sets $A$ and $B$ in $\rtwo$ we denote the the distance between them as $d(A,B) = \inf \LP \lvert \bfs - \bfs^\prime \rvert : \bfs \in A, \bfs^\prime \in B \RP$. For any random field $\LP T(\bfs) : \bfs \in \rtwo \RP$ we define the strong-mixing coefficient as 
\begin{equation}
\label{eq:MixingCoef}
\begin{split}
\alpha_T (u, v) = \sup \LP \tilde{\alpha}_T(A, B) : \; d(A,B) \geq u; \;\; \lvert A \rvert \leq v, \;\; \lvert B \rvert \leq v \RP,
\end{split}
\end{equation} where the supremum is taken over the set of all 2-dimensional rectangles $A$, $B$ in $\rtwo$, $u > 0$, $v > 0$; and 
$$
\tilde{\alpha}_T(A, B) = \sup \LP \lvert P(V_1 \cap V_2) - P(V_1) P(V_2) \rvert : V_1 \in \mathcal{F}_T(A), \; V_2 \in \mathcal{F}_T(B) \RP.
$$ For a distribution function $F(\cdot)$, we define the inverse of it as $F^{-1}(p)= \text{inf}\LP x: F(x) \geq p \RP$, for $p \in [0,1]$.

To definition of VS in Section~\ref{sec:vs} uses a square neighborhood $\mathcal{B}_{\delta}(\bfs_i)$ around the spatial point $\bfs_i$ with one side of $2\delta$ units. From this section we change the notation as: $\delta \equiv \delta_n$ and $\mathcal{B}_{\delta}(\cdot) \equiv \mathcal{B}_{\delta_n}(\cdot)$ to ensure that the size of the `local' neighborhood also varies with the sample size $n$. Let $M_{n(i)}^p$ be the smallest $p$-th sample quantile of $\LP w(\bfs_{i_1}), \dots , w(\bfs_{i_{n(i)}}) \RP$, the residuals associated with the observations in the square-neighborhood $\mathcal{B}_{\delta_n}(\bfs_i)$. We denote $\hat{F}_{n(i)}(x) = \LP \text{No. of $w(s_{i_j}) \leq x$}\RP/n(i)$, the empirical distribution function of the realizations from the $w$-process in the neighborhood around $\bfs_i$. Clearly, $M_{n(i)}^p = \hat{F}_{n(i)}^{-1}(p)$. By $\bar{F}_{n(i)}(x)$ we denote the distribution function defined as $\bar{F}_{n(i)}(x) = \Lp \sum_{j = 1}^{n(i)} F_{i_j}(x) \Rp/n(i)$, where $F_{i_j}$ is the distribution function of $w(\bfs_{i_j})$. The smallest $p$-th quantile of the distribution function $\bar{F}_{n(i)}$ is denoted by $\xi_{n(i)}^p$, i.e. $\xi_{n(i)}^p = \bar{F}_{n(i)}^{-1}(p)$.

\subsection{Consistency of the VS-based regression parameter estimator}
\label{subsec:AsympCons}
Here we focus on consistency of the $\hat{\bfbeta}_{\text{vs}}^{(m)}$, the Median-VS version of VS-based estimator. The conditions we need for that are following.
\begin{itemize}
\item[(C.1)] Number of observations in any unit square in $\mathcal{D}_n$ is in between $[ C_1  \frac{n}{\lambda_n^2} , C_2 \frac{n}{\lambda_n^2} ]$ for some positive constants $C_1$ and $C_2$ such that $C_1 < C_2$.

\item[(C.2)] Number of non-noisy observations ($Z(\bfs_i)$ such that $i \in G_n$) in any unit square of the sampling region is bounded below by $C_1 q_e n\lambda_n^{-2}$.

\item[(C.3)] The covariate process $\mathbf{x}(\bfs)$ is such that,
$\mathbf{x}(\bfs) = \mathbf{x}_0(\lambda_n^{-1} \bfs)$ where $\mathbf{x}_0: [0,1]^2 \to \rp$ is a differentiable function with bounded partial derivatives in $(0,1)^2$ and $\frac{1}{n} \sum_{i=1}^n \lVert \mathbf{x}(\bfs_i) \rVert^2  = O(1)$ for any set of $\LP \bfs_1, \dots , \bfs_n \RP \subset \mathcal{D}_n$ and for any $n \geq 2$.

\item[(C.4)] $\LP \epsilon(\bfs) : \; \bfs  \in \rtwo \RP$ is a second-order stationary random field such that, $\int_{\rtwo} \; \lvert \rho_\epsilon \Lp \bfh \Rp \rvert d\bfh  < \infty$; and for some $\kappa > 0$, $E \lvert\epsilon (\mathbf{0}) \rvert ^{2+\kappa} < \infty$ and $\alpha_{\epsilon}(u, v) \leq C u^{-\nu_1} v^{\nu_2}$ with $\nu_1 > 4/\kappa$ and $\nu_2 > 0$.

\item[(C.5)] $\delta_n^{-1}  + \lambda_n^{-1} \delta_n + n^{-1/2} \lambda_n \delta_n^{-1} \to 0.$

\item[(C.6)] $\underset{n}{\text{sup}} \underset{i \in \LP1, \dots , n \RP}{\text{sup}} \lvert \xi^p_{n(i)} \rvert < \infty$ for all $p \in \LP 0.25, 0.75 \RP$.

\item[(C.7)] For all $i \in \LP 1, \dots , n \RP$, $j \in \LP 1, \dots , n(i) \RP$ and $p \in \LP 0.25, 0.5, 0.75 \RP$, $F_{i_j}(\cdot)$ is absolutely continuous in an neighborhood of $\xi^p_{n(i)}$, i.e. $f_{i_j}(x) = \Lp d/dx \Rp F_{i_j}(x)$ exists in that neighborhood.

\item[(C.8)] For all $i \in \LP 1, \dots , n \RP,\;\;$ $0 < \underset{j \in \LP 1, \dots , n(i) \RP}{\text{inf}}f_{i_j}(\xi^p_{n(i)}) \leq \underset{j \in \LP 1, \dots , n(i) \RP}{\text{sup}}f_{i_j}(\xi^p_{n(i)}) < \infty$.

\item[(C.9)] $F_i^{-1}(0.5) = 0$ for all $i \in \LP 1, \dots , n \RP$.

\end{itemize}
In a later remark we have discussed how the conditions would change if we have considered the Mean-VS estimator, $\hat{\bfbeta}_{\text{vs}}^{(a)}$. We now comment briefly on the conditions. (C.1) assumes that the sampling design has to be such that the number of observations in any unit square varies as the same rate as of $n/\lambda_n^2$, the `average' number of observations per unit square. In (C.2) we restrict the number of noisy observations in any unit square by $C_1(1 - q_e)n\lambda_n^{-2}$ which combining with (C.1) specifies that at most $1 - q_e$ proportion of the observations in an unit block can be noisy, i.e. corresponding noise variances (denoted by $\tau_i^2$) are non-zero. (C.3) puts regularity conditions on the covariates; it states that the covariate process is an `inflated' version of a `smooth' process $\bfx_0(\bfs)$ in the prototype sampling region $[0,1]^2$, making $\bfx(\bfs)$ a slowly varying process over the space. The other condition in (C.3) is standard in regression theory which puts a bound on the magnitude of the covariates as compared to the sample size. (C.4) states the required moment and mixing conditions on the residual process $\LP \epsilon(\bfs) \RP$. Standard Gaussian process with `nicely' decaying covariance functions, e.g. Expoenential, Mat\'ern, Gaussian, etc. satisfies this condition. (C.5) makes sure that the size of the $\delta_n$-neighborhoods used to compute the VS increases with $n$ resulting in increasing number of observations for VS computation. But, at the same time the second term of condition (C.5) states that the block has to be small enough as compared to the whole sampling region retaining the `local' feature of VS. The third term in (C.5) states that the rate at which the samples are infilling in the $\delta_n$-neighborhood has to go to $\infty$ as $n \to \infty$. Conditions (C.6) - (C.8) are standard assumptions needed for quantile consistency of non-i.i.d. random variables (for details, see \citealt{sen68} and \citealt{jkg71}). (C.9) is needed to make the $w(\bfs_i)$'s marginal population median is equal to $0$.

We now state the consistency results of this paper. The following two propositions are the key to prove the rest of the results of our paper. In the following proposition we provide an asymptotic approximation of the Median-VS.
\begin{prop}
\label{prop:MedVSApprox}
Under conditions (C.1) to (C.9), for all $i \in \LP 1, 2, \dots , n \RP$,
$$
V^{(m)}(\bfs_i) = \exp\Lp - \frac{\lvert w(\bfs_i) \rvert}{\mathcal{I}_n(\bfs_i)} \Rp + O_p(a_n),
$$ where $\mathcal{I}_n(\bfs_i) = \xi_{n(i)}^{0.75} - \xi_{n(i)}^{0.25}$, i.e. the IQR of the distribution function $\bar{F}_{n(i)}(x)$ and $a_n$ is a sequence of positive reals $\downarrow 0$ given by,
$
a_n = n^{-1/2} \lambda_n \delta_n^{-1} + \delta_n \lambda_n^{-1}.
$
\end{prop}
The proof of this proposition is based on the extension of the Ghosh-Bahadur representation of sample quantiles for irregularly spaced spatial data (Lemma~\ref{lem:quantile-consist} in Appendix~\ref{app-subsec:aux-lem}). Before we comment on the implication of Proposition~\ref{prop:MedVSApprox}, let us state the following proposition.
\begin{prop}
\label{prop:IqrBound}
Under conditions (C.1) to (C.9), for all $i \in \LP 1, \dots , n \RP$
$$
\begin{aligned}
C_\epsilon^{(l)}(q_e) \leq \mathcal{I}_n(\bfs_i) = \xi_{n(i)}^{0.75} - \xi_{n(i)}^{0.25} \leq C_\epsilon^{(u)}(q_e)
\end{aligned}
$$ where, $C_\epsilon^{(l)}(q_e) = \sigma_\epsilon  \Lp F_\epsilon^{-1}(\text{max}\LP 1 - (0.25/q_e), 0 \RP) - F_\epsilon^{-1}(\text{min}\LP 0.25/q_e, 1 \RP) \Rp$ and\\  $C_\epsilon^{(u)}(q_e) = \sigma_\epsilon  \Lp F_\epsilon^{-1}(\text{min}\LP 0.75/q_e, 1 \RP) - F_\epsilon^{-1}(\text{max}\LP 1 - (0.75/q_e), 0 \RP) \Rp$.
\end{prop}
From Proposition~\ref{prop:IqrBound} we conclude that the scaling term, $\mathcal{I}_n(\bfs_i)$, in the Median-VS approximation in Proposition~\ref{prop:MedVSApprox} is bounded above by $C_\epsilon^{(u)}(q_e)$, which is a finite constant if $q_e > 0.75$, i.e. the proportion of noisy observations is less than $25\%$. Hence, under the assumption that $q_e > 0.75$, from Proposition~\ref{prop:MedVSApprox} we can see that if the noise variance associated with $w(\bfs_i)$ -- denoted by $\tau_i^2$ -- is large, with higher probability $\lvert w(\bfs_i) \rvert$ will take larger values making $\exp\Lp - \lvert w(\bfs_i) \rvert/ \mathcal{I}_n(\bfs_i) \Rp$ (from here denoted by $\tilde{V}^{(m)}(\bfs_i)$) -- which is bounded above by $\exp\Lp - \lvert w(\bfs_i) \rvert/ C^{(u)}_\epsilon(q_e) \Rp$ -- closer to $0$. Whereas, if $w(\bfs_i)$ has less variance then with higher probability $\lvert w(\bfs_i) \rvert$ will take values in a neighborhood of $0$ making $\tilde{V}^{(m)}(\bfs_i)$ closer to $1$. Also, note that the lower bound $C^{(l)}_\epsilon(q_e) > 0$ if we have $q_e > 0.5$. Hence, given that the poportion of good observations in the data is more than $75\%$, Proposition~\ref{prop:MedVSApprox} along with Proposition~\ref{prop:IqrBound} asymptotically justifies that Median-VS will give low scores to `bad' observations and high scores to `good' observations.

Next, we use the approximation in Proposition~\ref{prop:MedVSApprox} to come up with a representation of the VS-based mean parameter estimator $\hat{\bfbeta}^{(m)}_{\text{vs}}$ in the following theorem.
\begin{theo}
\label{theo:VS-Mean-Prepresentation}
Under conditions (C.1) to (C.9),
\begin{equation}
\label{eq:MedVSRepTheo}
    \begin{split}
        \hat{\bfbeta}_{\text{vs}}^{(m)} = \bfbeta + \Lp \frac{1}{n} \bfX^\prime E\Lp\tilde{\mathbf{D}}_v\Rp \bfX \Rp^{-1} \Lp \frac{1}{n} \bfX^\prime \tilde{\mathbf{D}}_v \mathbf{w} \Rp + O_p \Lp a_n \Rp,
    \end{split}
\end{equation}
where $\tilde{\mathbf{D}}_v = \text{diag}\Lp \tilde{V}^{(m)}(\bfs_1), \dots , \tilde{V}^{(m)}(\bfs_n) \Rp$ and $\mathbf{w} = \Lp w(\bfs_1), \dots , w(\bfs_n) \Rp^\prime$.
\end{theo}
The proof of Theorem~\ref{theo:VS-Mean-Prepresentation} is given in Appendix~\ref{app-subsubsec:med-vs-reg-rep}. Now, we add some remarks regarding the representation of Median-VS-based estimator given in Equation~\ref{eq:MedVSRepTheo}.
\begin{rem}
\label{rem:Med-VS-representation}
The representation in Equation~\ref{eq:MedVSRepTheo} asymptotically approximates the estimation error in $\hat{\bfbeta}^{(m)}_{\text{vs}}$ by the random variable given by $\Lp \frac{1}{n} \bfX^\prime E\Lp\tilde{D}_v\Rp \bfX \Rp^{-1} \Lp \frac{1}{n} \bfX^\prime \tilde{D}_v \mathbf{w} \Rp.$ Observe that, if $w(\bfs_i)$ are marginally symmetric around $0$, then $E\Lp \tilde{V}(\bfs_i) w(\bfs_i) \Rp = 0$. Moreover, as proved in Section~\ref{app-subsubsec:vs-ineq}, $E\Lp \tilde{V}(\bfs_i)^2 w(\bfs_i)^2 \Rp \leq \exp(-2)(C_\epsilon^{(u)}(q_e))^2$. Using this, in the next section we show that the MSE in VS-based estimation of the regression parameter is asymptotically not dependent on the high noise variances, i.e. $\tau_i^2$. On the contrary, the same is not true for the OLS estimator as, $E\Lp w(\bfs_i)^2 \Rp = \sigma_\epsilon^2 + \tau_i^2$. This observation highlights the importance of the representation in Equation~\ref{eq:MedVSRepTheo}. We elaborately discuss the advantage of VS-based estimation in this regard in Section~\ref{subsec:AsympeffVSBased}.
\end{rem}
 We use Theorem~\ref{theo:VS-Mean-Prepresentation} to prove the consistency of $\hat{\bfbeta}^{(m)}_{\text{vs}}$ in the following corollary. For that we need the a set of additional conditions as follows.
\begin{itemize}
    \item[(C.10)] $\frac{1}{n}\bfX^\prime \bfX \to \bfC_X \succ 0$.
    \item[(C.11)] $E\Lp \tilde{D}_v \mathbf{w} \Rp = \mathbf{0}$.
    \item[(C.12)] The proportion of non-noisy observations, i.e. $q_e$ is strictly greater than $0.75$.
    \item[(C.13)] For any real number $a > 0$, $\psi_\epsilon(a) = \int_0^\infty e^{-x} P \Lp  \lvert \epsilon(\bfs) \rvert < a x \Rp dx > 0$. 
\end{itemize}

We now briefly comment on the above conditions. Condition (C.10) is standard assumption needed for the consistency of the least squares based mean parameter estimators. (C.11) is trivially valid if the $w(\bfs_i)$'s are marginally symmetric around $0$. (C.12) is needed as to get a finite upper-bound on the scaling component of $\tilde{V}^{(m)}(\bfs_i)$, i.e. $I_n(\bfs_i)$, we need the proportion of noisy observations to be less than $0.25$. (C.13) is required to ensure the invertibility of the matrix $\Lp n^{-1} \bfX^\prime E\Lp\tilde{\mathbf{D}}_v\Rp \bfX \Rp$. This condition is trivially true when the marginal distribution of $\epsilon(\bfs)$ has positive mass around a neighborhood of its center.
\begin{cor}
\label{cor:MedVSConsistent}
Under conditions (C.1) to (C.13),
$
\hat{\bfbeta}_{\text{vs}}^{(m)} = \bfbeta + O_p \Lp a_n \Rp.
$
\end{cor}
Corollary~\ref{cor:MedVSConsistent} not only proves the consistency of the Median-VS-based estimator, but also provides an order of rate of convergence in probability for the Median-VS-based estimator. We state the proof of Corollary~\ref{cor:MedVSConsistent} in Appendix~\ref{app-subsubsec:med-vs-consist-cor}. We now put some remarks on the consistency of the Mean-VS-based estimator.
\begin{rem}
\label{rem:mean-vs-consistency}
Similar consistency results of Mean-VS-based estimator $\Lp \hat{\bfbeta}_{\text{vs}}^{(a)}\Rp$ can also be proved but that requires a set of different conditions than that are stated in (C.1) - (C.9). As the sample standard deviation is used in the definition of Mean-VS we need $\frac{1}{n} \sum_{i=1}^n \lVert \mathbf{x}(\bfs_i) \rVert^4  = O(1)$ as well as all the error components including $\epsilon(\bfs)$ must have finite fourth order moments. In addition, the mixing condition in (C.4) changes as we would require $\epsilon$-process to be fourth-order stationary and all the correlations of the form $\text{cor}\Lp \epsilon^a(\cdot),  \epsilon^b(\cdot + \bfh) \Rp$, for all $a,b \in \LP 1, \dots , 4 \RP$, have to be integrable over the space. Conditions  (C.5) - (C.9) are not needed as those are specific to the existence of the population quantiles used in the results of the Median-VS-based estimator.
\end{rem}

Though the consistency of the VS-based estimators has been proved in this section, that does not necessarily point out the advantages of the VS-based estimation over the standard methods, e.g. the OLS, when it comes to analyzing noisy spatial data. In the next subsection, we investigate the asymptotic mean squared errors of the VS-based as well as the OLS estimator. For the next section we only consider the Median-VS-based regression estimator ($\hat{\bfbeta}^{(m)}_{\text{vs}}$) and hence, for simplicity of nations, we denote it by $\hat{\bfbeta}_{\text{vs}}$.

\subsection{Asymptotic efficiency of VS-based estimators}
\label{subsec:AsympeffVSBased}
From Theorem~\ref{theo:VS-Mean-Prepresentation}, we can write $\hat{\bfbeta}_{\text{vs}} - \bfbeta = \mathbf{l}_n^{\text{vs}} + O_p(a_n)$, where the leading error term in the VS-based estimation is given by, $\mathbf{l}_n^{\text{vs}} = \Lp n^{-1} \bfX^\prime E\Lp\tilde{\mathbf{D}}_v\Rp \bfX \Rp^{-1} \Lp n^{-1} \bfX^\prime \tilde{\mathbf{D}}_v \mathbf{w} \Rp$. On the other hand, due to linearity of the OLS estimator, $\hat{\bfbeta}_{\text{ols}} - \bfbeta = \mathbf{l}_n^{\text{ols}}$, where it can be easily shown that $\mathbf{l}_n^{\text{ols}} = \Lp n^{-1} \bfX^\prime \bfX \Rp^{-1} \Lp n^{-1} \bfX^\prime \mathbf{w} \Rp$. Before stating the results of this section, we need to introduce a few more notations. It can be shown easily that $E\Lp \lVert \mathbf{l}_n^{\text{ols}}  \rVert^2 \Rp = n^{-1} \text{tr}\Lp \bfH_n \boldsymbol{\Sigma}_w \Rp$ (details are given in Appendix~\ref{app-subsubsec:ols-ineq}), where $\bfH_n = n^{-1}\mathbf{X}(n^{-1}\mathbf{X}^\prime \mathbf{X})^{-2}\mathbf{X}^\prime$, $\boldsymbol{\Sigma}_w = \var\Lp \mathbf{w} \Rp$. For any symmetric matrix $\bfA$, by $\lambda_{\text{min}}\Lp \bfA \Rp$, we denote the minimum eigenvalue of $\bfA$. Let us introduce the following notations: $M_X = \underset{\bfs \in \mathcal{D}_n}{\text{sup}} \underset{1 \leq j \leq p}{\text{max}} \; x^2_j(\bfs) $; $C^{(0)} = 2^{-1}\text{tr}\Lp \bfC_X^{-1} \Rp$; $C^{(1)} = 2 p \Lp\lambda_{\text{min}}\Lp \bfC_X \Rp\Rp^{-2} M_X$; and finally $\mathcal{M}_n(\varepsilon) = \LP 1 \leq i \leq n : \Lp \bfH_n \Rp_{ii} > n^{-1} \varepsilon \RP$, where by $\Lp \bfA \Rp_{ij}$ we denote the $ij$-th element of the matrix $\bfA$. Now, we have the required notation to state the main results of this section.
\begin{theo}
\label{theo:vs_ineq}
Let assumptions (C.1) to (C.13) hold true. Then, for positive constants $C_1(q_e)$, $C_2$ and $C_3$,
\begin{equation}
\label{eq:vs_ineq}
    \begin{split}
        E\Lp \lVert \mathbf{l}_n^{\text{vs}} \rVert^2\Rp \leq C_3 \cdot \Lp n^{-1}\Lp C_1(q_e) \Rp^{2} + C_2\lambda_n^{-4}\Rp,
    \end{split}
\end{equation} where $C_1(q_e) = \Lp C_\epsilon^{(u)}(q_e)/\psi_\epsilon\Lp C_\epsilon^{(l)}(q_e) \Rp\Rp$ and $C_2$, $C_3$ are independent of the noise variance parameters ($\tau_i^2$ for $i \in G_n$, and $q_e$).
\end{theo}
\begin{theo}
\label{theo:ols_ineq}
Let assumptions (C.1) to (C.4) and (C.10) hold true. Then, for any $0 < \varepsilon < \text{min}\Lp C^{(0)}, C^{(1)} \Rp$ and large enough $n$,
\begin{equation}
\label{eq:ols_ineq}
    \begin{split}
        E\Lp \lVert \mathbf{l}_n^{\text{ols}} \rVert^2 \Rp > n^{-1}\varepsilon\Lp \sigma_\epsilon^2 \Lp C^{(0)} - \varepsilon \Rp\Lp C^{(1)} - \varepsilon \Rp^{-1} + n^{-1}\sum_{i \in \mathcal{M}_n(\varepsilon)\cap G_n^c} \tau_i^2 \Rp - C_1 \lambda_n^{-4},
    \end{split}
\end{equation}
with $\lvert \mathcal{M}_n(\varepsilon) \rvert \geq \frac{C^{(0)} - \varepsilon}{C^{(1)} - \varepsilon}n$. Here $C_1$ is a positive constant not dependent on the error variance parameters ($\tau_i^2$ for $i \in G_n$, and $q_e$).
\end{theo}
\begin{cor}
\label{cor:ols_ineq}
Let assumptions (C.1) - (C.4) and (C.10) hold true. In addition, assume $\text{inf}_n \; \underset{1\leq i \leq n}{\text{min}} \lVert \bfX[i,] \rVert^2 > 0$. Then, there exists an $\varepsilon > 0$ such that for large enough $n$,
\begin{equation}
\label{eq:ols_ineq_cor}
    \begin{split}
        E\Lp \lVert \mathbf{l}_n^{\text{ols}} \rVert^2 \Rp > n^{-1}\Lp \varepsilon\sigma_\epsilon^2  + n^{-1}\sum_{i \in G_n^c} \tau_i^2 \Rp - C_1 \lambda_n^{-4},
    \end{split}
\end{equation} for some positive constant $C_1$ not dependent on the noise model parameters.
\end{cor}
The proofs of Theorem~\ref{theo:vs_ineq}, Theorem~\ref{theo:ols_ineq} and Corollary~\ref{cor:ols_ineq} are stated in Appendix~\ref{app-subsubsec:vs-ineq}, \ref{app-subsubsec:ols-ineq} and \ref{app-subsubsec:ols-ineq-cor} respectively.  Next, we put some remarks on the above results.
\begin{rem}
From Theorem~\ref{theo:vs_ineq} we see that the MSE of the leading error term in the Median-VS-based regression estimation can be bounded above by some constant $C_3$ times $n^{-1}\Lp C_1(q_e) \Rp^{2} + C_2\lambda_n^{-4}$, which goes to $0$ as $n \to \infty$, and is dependent on the noise model parameters only through the proportion of the good observations $q_e$. Clearly, the MSE of the Median-VS-based regression estimator is not dependent on the noise variances (denoted by $\tau_i^2$) associated with the `bad' observations. Note that, $\psi_\epsilon(\cdot)$ is an increasing function. We have already mentioned that as $q_e$ increase, $C_\epsilon^{(u)}(q_e)$ decreases and $C_\epsilon^{(l)}(q_e)$ increases, i.e. $C_1(q_e)$ is a decreasing function of $q_e$. Hence, if the proportion of good observations increases, the upper bound in Equation~\ref{eq:vs_ineq} decreases, restricting the range of $E\Lp \lVert \mathbf{l}_n^{\text{vs}} \rVert^2\Rp$ from above. The other term, $C_2\lambda_n^{-4}$ has hardly any effect on $E\Lp \lVert \mathbf{l}_n^{\text{vs}} \rVert^2\Rp$ as $\lambda_n^{-4}$ is of much smaller order as compared to $n^{-1}$.
\end{rem}
\begin{rem}
The lower-bound for the MSE of the OLS estimator, as we can see in Theorem~\ref{theo:ols_ineq} and Corollary~\ref{cor:ols_ineq}, is dependent on the additional noise variances of the corrupted observations through the summation $n^{-1}\sum_{i \in G_n^c \cap \mathcal{M}_n(\varepsilon)} \tau_i^2$ and $n^{-1}\sum_{i \in G_n^c} \tau_i^2$ respectively. Clearly, if the noise variances corresponding to some of the observations are high, efficiency of the OLS estimator will be significantly affected. For example, consider the case where $\tau_i^2 \approx n^c$ for some constant $c > 0$. Then for a given $q_e$, the lower-bound in Equation~\ref{eq:ols_ineq_cor}, is \textit{of order} $n^{c-1}$. In Table~\ref{tab:RECompEx} we show the advantage of using VS-based estimator as opposed to the OLS one in terms of the asymptotic order of relative efficiency.
\begin{table}[h]
\centering
\caption{\small Comparison of orders of MSEs of OLS and VS under the example case: $\tau_i^2 \approx n^c$.}
\label{tab:RECompEx}
\resizebox{0.32\columnwidth}{.12\textwidth}{%
\begin{tabular}{|cc|cc|c|}
 \toprule
 \textbf{n} & \textbf{c} & \textbf{OLS-LB} & \textbf{VS-UB} & \textbf{RE-LB} \\ 
  \hline
\multirow{3}{*}{100} & 0.1 & 0.016 & 0.010 & 1.585 \\ 
 & 0.5 & 0.100 & 0.010 & 10.000 \\ 
 & 0.8 & 0.398 & 0.010 & 39.811 \\ \hline
\multirow{3}{*}{500} & 0.1 & 0.004 & 0.002 & 1.862 \\ 
 & 0.5 & 0.045 & 0.002 & 22.361 \\ 
 & 0.8 & 0.289 & 0.002 & 144.270 \\ \hline
\multirow{3}{*}{1000} & 0.1 & 0.002 & 0.001 & 1.995 \\ 
 & 0.5 & 0.032 & 0.001 & 31.623 \\ 
 & 0.8 & 0.251 & 0.001 & 251.189 \\ 
   \hline
\end{tabular}%
}
\end{table}
Here, we refer the order of the lower bound of the MSE of the OLS estimator by OLS-LB, the upper bound for the MSE of the VS-based estimator by VS-UB and, the lower bound for the relative efficiency of VS-based estimator to the OLS one (defines as MSE(VS)/MSE(OLS)) by RE-LB. Table~\ref{tab:RECompEx} clearly portrays the advantage of the VS-based estimation as compared to the OLS one: if the order of the noise variances associated with even a `small' portion of the observations is `high' , due to the robustness of the VS-based estimator against the noisy observations, there is a significant gain in efficiency relative to the OLS. 
\end{rem}


\section{Simulation Study}
\label{sec:SimStudy}
In this section, we summarize our efforts to evaluate the VS-based estimator numerically. We first describe the simulation designs and then, we state the results and draw inference from them.

\subsection{Simulation Setup}
\label{subsec:SimSetup}
We use the following spatial regression model to simulate the realizations from the the `true' random field $\LP Y(\bfs) \RP$:
\begin{equation}
\label{eq:SimMod_LM_Y}
    \begin{split}
        Y(\bfs_i) = \beta_0 + \Lp \beta_x, \beta_y \Rp^\prime \bfs_i + \beta_h \; h(\bfs_i) + \epsilon(\bfs_i), \quad \text{for} \;\; i \in \LP 1, \dots , n \RP,
    \end{split}
\end{equation} where, $\bfbeta = \Lp \beta_0, \beta_x, \beta_y, \beta_h \Rp^\prime$ is the vector of regression parameters; $h(\bfs)$ is a deterministic function of location $\bfs$; and $\LP \epsilon(\bfs) \RP$ is a mean zero second-order stationary spatially correlated process. To define the function $h(\bfs)$ over the sampling region, we use the deterministic function $h(\bfs) = H_1 \cdot \sum_{j=1}^{H_2} w_h(j) f(\bfs; \; \boldsymbol{\mu}_j , \Sigma_j) + H_3$, where $f(\cdot; \boldsymbol{\mu}, \Sigma)$ denotes the bivariate normal density with mean $\boldsymbol{\mu}$ and covariance matrix $\Sigma$ and $\LP \Lp \boldsymbol{\mu}_j, \Sigma_j \Rp : 1 \leq j \leq H_2 \RP$ is a fixed set of vectors and matrices. The choice of this function is motivated from the elevation function used in the simulations of \cite{chak}. The residual vector $\Lp \epsilon(\bfs_1), \dots , \epsilon(\bfs_n) \Rp^\prime$ are sampled from a mean zero Gaussian process with the Exponential covariance function given as $
C^{\text{exp}}(\bfh;  \boldsymbol{\theta}) = \sigma_\epsilon^2 \exp \Lp \frac{\lVert \bfh \rVert}{\rho} \Rp  + \tau^2 \mathbb{I}(d = 0)$ where, $\mathbb{I}(\cdot)$ denotes the indicator function. The covariance parameter vector of interest is $\boldsymbol{\theta} = \Lp \tau^2, \sigma_\epsilon^2, \rho \Rp^\prime$, where $\tau^2$ is the nugget effect, $\sigma_\epsilon^2, \rho$ are the partial sill and range parameters respectively (for details see \citealt{haskard07}; page 37 \citealt{gelfand10}).

To include noise in the varying-quality observations $\Lp Z(\bfs_1), \dots  , Z(\bfs_n) \Rp^\prime$ using the additive-multiplicative noise structure given in Equation~\ref{eq:AddMultNoiseMod}, we use the following distributions for the additive and multiplicative components. Given a set of good observations $G_n$, for $i \in G_n$, the $\epsilon_{M_i}$ and $\epsilon_{A_i}$ are identically equal to $1$ and $0$, respectively. For $i \in G_n^c$, we take
\begin{equation}
\label{eq:error_dist_sim}
\begin{split}
\epsilon_{M_i} \underset{\text{indep.}}{\sim} 2 \times \text{Beta}(\alpha_{M},& \alpha_{M}), \quad \epsilon_{A_i} \underset{\text{indep.}}{\sim} N(0,\sigma_{A}^2).
\end{split}
\end{equation} Here, the variance associated with the multiplicative component of a noisy observation is given by $\sigma_{M}^2 = \frac{1}{2\alpha_{M} + 1}$. To select the set of `good' observations, i.e. $G_n$, we have arbitrarily selected $\lfloor q_e n \rfloor$ many (fixed) indices among $\LP 1, \dots , n \RP$ prior the simulations. Note that, the choice of multiplicative error distribution in \ref{eq:error_dist_sim} restricts its realizations to be in $[0,2]$, and also, ensures that the multiplicative errors are symmetric around 1.

Estimation of the regression parameters through VS-based and the OLS method do not require the covariance parameters to be estimated. Whereas, the GLS estimator uses the covariance information which is often unknown in real data analysis. Hence, the covariance parameters are estimated from the de-trended spatial observations and then the estimated covariance parameters are used to compute GLS estimator. Then the updated mean estimates are used to compute the covariance parameters and we continue until the parameter estimates converge. This estimator is called the estimated GLS (referred as EGLS, for details see page 23 \citealt{cressie93}). In the following section, we compare the performance of VS-based (Mean-VS, Median-VS) and commonly used least squares based (OLS and EGLS) techniques for estimation of the regression parameters.

Though the theoretical analysis of the VS-based covariance parameter estimation is beyond the scope of this paper, in our simulation studies we evaluate the accuracy of the same to capture the dependence structure of the spatial process. The reasons are following. First, the boxplots of the regression parameter estimates in Section B.1 in the supplementary material suggest that the robust approach (robust regression with Huber's loss, \citealt{huber81}) has similar performance as compared to the VS-based methods. But, extension of the robust approach for estimation of the covariance parameters is not straightforward. Whereas, the VS computed from the data can be incorporated in the covariance analysis to reduce the effects of `bad' observations in the estimation. Hence, comparison of the VS-based technique of covariance estimation needs to be evaluated as compared to the other robust techniques of covariance parameter estimations. We consider the following two methods in that regard: the robust WLS based variogram model fitting by \cite{cressie80} (referred as WLS in Table~\ref{tab:CovParamMSE}) and the robust-REML technique proposed by \cite{kunsch11} and implemented by \cite{georob18}. Second, we want to justify the use of VS-based covariance analysis in the real data example in Section~\ref{sec:readl-dat-ex}.

For all the simulation results in this section we take $\bfbeta = \Lp 70, 5, -2, -0.05 \Rp^\prime$ and $\bftheta = \Lp 0, 6, 1\Rp^\prime$. To investigate the performances of the different types of estimators with varying noise level, we consider the following set of values of the noise parameters: (a) $\sigma_A \in \LP 5, 50, 100 \RP$, (b) $\alpha_M \in \LP 2, 0.5, 0.05 \RP$, and (c) $ q_e \in \LP 0.95, 0.85, 0.75 \RP$. The set of values for each of the above parameters are written in the order of increasing noise variance and proportion. For example, if the first values of all the three sets are considered then the additional noise variance of a corrupted observation at location $\bfs$ is $0.2\Lp \bfx(\bfs)^\prime \bfbeta \Rp^2 + 26.2$, and the proportion of such observations is $5\%$; whereas the same variance will be $0.91\Lp \bfx(\bfs)^\prime \bfbeta \Rp^2 + 10005.46$ if the last values of the sets are considered and the proportion of such noisy observations will be $25\%$. In the tables in the next section, when one of the noise model parameters are varying, the other parameters are fixed at the lowest noise level, i.e. if $\sigma_A$ is varying, then $\alpha_M$ is fixed at $2$ and $q_e$ is fixed at $0.95$.

\subsection{Simulation Results}
\label{subsec:SimResult}
Table~\ref{tab:RegParamMSE} shows the empirical mean square errors of the considered mean parameter estimators. In the final column of Table~\ref{tab:RegParamMSE}, the relative efficiency (RE) of the Median-VS-based regression estimator with respect to the OLS one is reported to emphasize the superiority of VS-based estimation in the analysis of noisy spatial data.
\begin{table}[h]
\centering
\caption{\small Performance of regression parameter estimators under additive-multiplicative noise model.}
\label{tab:RegParamMSE}
\resizebox{0.72\columnwidth}{.43\textwidth}{%
\begin{tabular}{|cc|cc|cc|c|}
\toprule
  & \multicolumn{1}{c|}{} &
  \multicolumn{2}{c|}{\textbf{VS-based MSE}} &
  \multicolumn{2}{c|}{\textbf{LS-based MSE}} &
  \textbf{RE(Med-VS, OLS)=}\\
  \cline{3-6}
  
 \textbf{$\sigma_A$} & $\mathbf{n}$ &
  \textbf{Med-VS} &
  \textbf{Mean-VS} &
  \textbf{OLS} & \textbf{EGLS} &  \textbf{MSE(OLS)/MSE(Med-VS)} \\ 
  \hline
  \hline
\multirow{3}{*}{5} &  500 & 1.807 & 1.959 & 3.237 & 5.620 & 1.791 \\ 
 & 1000 & 1.166 & 1.184 & 1.662 & 3.416 & 1.425 \\ 
 & 5000 & 0.567 & 0.574 & 0.746 & 1.436 & 1.317 \\ \hline
\multirow{3}{*}{50} &  500 & 1.806 & 1.906 & 4.783 & 14.862 & 2.648 \\ 
 & 1000 & 1.134 & 1.165 & 2.534 & 6.112 & 2.234 \\ 
 & 5000 & 0.701 & 0.693 & 1.055 & 2.114 & 1.505 \\ \hline
\multirow{3}{*}{100} &  500 & 2.350 & 2.905 & 12.938 & 36.208 & 5.505 \\ 
 & 1000 & 0.969 & 1.102 & 7.817 & 13.895 & 8.067 \\ 
 & 5000 & 0.618 & 0.595 & 1.618 & 6.064 & 2.619 \\ \hline
 \toprule
  & \multicolumn{1}{c|}{} &
  \multicolumn{2}{c|}{\textbf{VS-based MSE}} &
  \multicolumn{2}{c|}{\textbf{LS-based MSE}} &
  \textbf{R.E.(Med-VS, OLS)=}\\
  \cline{3-6}
  
 \textbf{$\sigma_M$} & $\mathbf{n}$ &
  \textbf{Med-VS} &
  \textbf{Mean-VS} &
  \textbf{OLS} & \textbf{EGLS} &  \textbf{MSE(OLS)/MSE(Med-VS)} \\ 
  \hline
  \hline
\multirow{3}{*}{0.447} &  500 & 1.889 & 1.945 & 2.771 & 4.481 & 1.467 \\ 
 & 1000 & 1.005 & 1.033 & 1.559 & 2.942 & 1.552 \\ 
 & 5000 & 0.572 & 0.578 & 0.699 & 1.352 & 1.220 \\\hline 
\multirow{3}{*}{0.707} &  500 & 2.219 & 2.253 & 4.098 & 13.248 & 1.847 \\ 
 & 1000 & 1.063 & 1.117 & 2.685 & 6.031 & 2.525 \\ 
 & 5000 & 0.647 & 0.641 & 0.919 & 2.527 & 1.420 \\ \hline
\multirow{3}{*}{0.953} &  500 & 2.221 & 2.422 & 6.182 & 20.996 & 2.784 \\ 
 & 1000 & 1.201 & 1.206 & 3.578 & 9.048 & 2.978 \\ 
 & 5000 & 0.669 & 0.671 & 1.476 & 4.746 & 2.206 \\ \hline
 \toprule
  & \multicolumn{1}{c|}{} &
  \multicolumn{2}{c|}{\textbf{VS-based MSE}} &
  \multicolumn{2}{c|}{\textbf{LS-based MSE}} &
  \textbf{R.E.(Med-VS, OLS)=}\\
  \cline{3-6}
  
 \textbf{$q_e$} & $\mathbf{n}$ &
  \textbf{Med-VS} &
  \textbf{Mean-VS} &
  \textbf{OLS} & \textbf{EGLS} &  \textbf{MSE(OLS)/MSE(Med-VS)} \\ 
  \hline
  \hline
\multirow{3}{*}{0.95} &  500 & 2.121 & 2.160 & 3.080 & 5.697 & 1.452 \\ 
& 1000 & 1.121 & 1.110 & 1.447 & 2.527 & 1.291 \\ 
& 5000 & 0.742 & 0.741 & 0.836 & 1.281 & 1.126 \\ \hline
\multirow{3}{*}{0.9} &  500 & 1.783 & 2.026 & 4.274 & 8.825 & 2.397 \\ 
& 1000 & 1.129 & 1.177 & 2.362 & 5.475 & 2.092 \\ 
& 5000 & 0.634 & 0.635 & 0.833 & 2.288 & 1.314 \\ \hline
\multirow{3}{*}{0.8} &  500 & 2.017 & 2.351 & 6.074 & 15.545 & 3.011 \\ 
& 1000 & 1.087 & 1.267 & 3.567 & 9.290 & 3.283 \\ 
& 5000 & 0.550 & 0.565 & 1.105 & 4.618 & 2.010 \\ \hline
 \toprule
   \end{tabular}%
}
\end{table}
It is evident from Table~\ref{tab:RegParamMSE} that the VS-based estimation has better performance than the standard OLS and EGLS methods across all the cases. The increasing noise variance has very little effect on the performance of the VS-based estimation in terms of MSE. For instance, consider the case when $n = 500$. If the standard deviation of the associated additive noise increases from $5$ to $50$, the MSE of Median-VS-based estimator has hardly changed, whereas the MSE of the EGLS and OLS estimators have increased by approximately $164.4\%$ and $47.8\%$, in the respective order. The consistency of the VS-based estimator is also prominent from Table~\ref{tab:RegParamMSE} -- in each of the scenarios, as sample size increases, the MSE is going close to $0$. The relative efficiency column justifies the superiority of the VS-based estimator to the least squares-based estimators. For example, when $\sigma_A = 100$ and the sample size $n = 1000$, i.e. very high noise variance is associated with $5\%$ of the observations -- to achieve same accuracy as of the Median-VS-based estimators the OLS and the EGLS methods need approximately $8$ and $14.5$ times more observations. Also, given a sample size, the relative efficiency increases if the magnitude of the noise variances associated with the `bad' observations increase (controlled by $\sigma_M$ and $\sigma_A$) or the proportion of the `bad' observations increases. Note that, the advantage of the VS-based estimation for varying proportion of the non-noisy data (i.e. $q_e$) is not as `good' as in the case of varying additive noise variance $\sigma_A$. The reason is that the effect of the increasing noise variances on the MSE of the VS-based estimators is much less as cmpared to the increase of proportion of noisy observations, which we have established theoretically in Section~\ref{subsec:AsympeffVSBased}. 

Observe that, in Table~\ref{tab:RegParamMSE}, the accuracy is hampered severely if instead of the OLS estimator EGLS is used to estimate the regression parameter. One of the possible reasons behind this is the inefficient estimation of the covariance parameters using robust weighted least squares technique.
\setlength{\tabcolsep}{12pt}
\begin{table}[h]
\centering
\caption{\small Performance of covaraince parameter estimators under additive-multiplicative noise model.}
\label{tab:CovParamMSE}
\resizebox{0.60\columnwidth}{.38\textwidth}{%
\begin{tabular}{|cc|cc|cc|}
\toprule
  & \multicolumn{1}{c|}{} &
  \multicolumn{2}{c|}{\textbf{psill MSE}} &
  \multicolumn{2}{c|}{\textbf{range MSE}}\\
  \cline{3-6}
  
 \textbf{$\sigma_A$} & $\mathbf{n}$ &
  \textbf{VS} &
  \textbf{WLS} &
  \textbf{VS} & \textbf{WLS}\\ 
  \hline
  \hline
\multirow{3}{*}{5} &  500 & 19.707 & 2058.760 & 2.471 & 3.099 \\ 
  & 1000 & 1.255 & 1992.875 & 0.123 & 2.697 \\ 
  & 5000 & 0.532 & 609.690 & 0.032 & 17.537 \\  \hline
\multirow{3}{*}{50} & 500 & 36.275 & 1334.724 & 4.549 & 0.580 \\ 
  & 1000 & 1.800 & 1820.486 & 0.146 & 0.639 \\ 
  & 5000 & 0.757 & 1163.299 & 0.046 & 11.401 \\   \hline
\multirow{3}{*}{100} & 500 & 45.653 & 682.913 & 9.477 & 0.502 \\ 
  & 1000 & 1.675 & 913.053 & 0.186 & 0.506 \\ 
  & 5000 & 0.933 & 1802.586 & 0.060 & 10.526 \\ \hline
 \toprule
  & \multicolumn{1}{c|}{} &
  \multicolumn{2}{c|}{\textbf{psill MSE}} &
  \multicolumn{2}{c|}{\textbf{range MSE}}\\
  \cline{3-6}
  
 \textbf{$\sigma_M$} & $\mathbf{n}$ &
  \textbf{VS} &
  \textbf{WLS} &
  \textbf{VS} & \textbf{WLS}\\ 
  \hline
  \hline
\multirow{3}{*}{0.447} & 500 & 7.712 & 1996.900 & 0.928 & 0.757 \\ 
& 1000 & 1.401 & 1863.389 & 0.164 & 5.779 \\ 
& 5000 & 0.580 & 577.577 & 0.047 & 6.312 \\  \hline 
\multirow{3}{*}{0.707} & 500 & 14.324 & 2045.811 & 2.097 & 13.983 \\ 
& 1000 & 1.663 & 1566.315 & 0.170 & 4.272 \\ 
& 5000 & 0.820 & 792.043 & 0.038 & 8.928 \\  \hline
\multirow{3}{*}{0.953} & 500 & 28.337 & 1578.990 & 3.621 & 54.341 \\ 
& 1000 & 1.507 & 1537.173 & 0.201 & 16.551 \\ 
& 5000 & 1.134 & 1185.805 & 0.091 & 1.645 \\   \hline
 \toprule
  & \multicolumn{1}{c|}{} &
  \multicolumn{2}{c|}{\textbf{psill MSE}} &
  \multicolumn{2}{c|}{\textbf{range MSE}}\\
  \cline{3-6}
  
 \textbf{$q_e$} & $\mathbf{n}$ &
  \textbf{VS} &
  \textbf{WLS} &
  \textbf{VS} & \textbf{WLS}\\ 
  \hline
  \hline
\multirow{3}{*}{0.95} &  500 & 11.786 & 1803.304 & 1.302 & 2.062 \\ 
& 1000 & 1.230 & 2385.300 & 0.146 & 2.884 \\ 
& 5000 & 0.665 & 531.284 & 0.047 & 12.393 \\  \hline
\multirow{3}{*}{0.9} &  500 & 5.216 & 2318.057 & 0.906 & 5.762 \\ 
& 1000 & 1.292 & 1535.030 & 0.098 & 5.707 \\ 
& 5000 & 1.063 & 817.137 & 0.026 & 2.803 \\  \hline
\multirow{3}{*}{0.8} &  500 & 36.674 & 1745.520 & 1.865 & 0.658 \\ 
& 1000 & 6.300 & 1308.637 & 0.116 & 13.993 \\ 
& 5000 & 4.584 & 1428.831 & 0.049 & 0.790 \\  \hline
 \toprule
   \end{tabular}%
}
\end{table}
Hence, in Table~\ref{tab:CovParamMSE}, we have investigated the performance of the VS-based covariance estimation and compared it with that of the standard approach: robust weighted least squares (WLS) based variogram estimator. Clearly, the WLS approach has failed completely to estimate the covariance parameters -- partial sill (or psill, denoted by $\sigma_\epsilon^2$) and range (denoted by $\rho$). Due to the added noise variance in the data the optimization in the variogram fitting in WLS method is not stable. On the other hand, VS-based variogram estimation has reasonably good accuracy, given that the proportion of the noisy observations are not too large. Also, in most of the cases, if the noise model parameters are held fixed, the MSE of the VS-based estimators decreases, as the sample size $n$ increase, which justifies the consistency of the VS-based variogram model fitting technique. But, in the case where the proportion of noisy observations is large (e.g. $q_e = 0.8$, i.e. $20\%$ of the observations are noisy) the accuracy of the VS-based estimation is hampered. Clearly, the accuracy of the VS-based covariance estimation is more sensitive to the proportion of noisy observations in the data, rather than the magnitude of the noise variances -- the same fact that we have established theoretically for the VS-based mean estimation in Section~\ref{subsec:AsympeffVSBased}.

Next, in Table~\ref{tab:VSvsGR}, we briefly report the results of our comparative analysis of VS-based technique with another robust covariance estimation method in geostatistics -- robust REML proposed by \cite{kunsch11} and implemented in R-package \texttt{georob} by \cite{georob18} (referred as GR in Table~\ref{tab:VSvsGR}).
\begin{table}[h]
\caption{Comparison of VS-based cov. estimation with GR}\label{tab:VSvsGR}
\centering
\resizebox{.63\columnwidth}{.11\textwidth}{%
\begin{tabular}{|cc|cccc|}
\toprule
$\mathbf{\sigma_A}$ & $\mathbf{n}$ & \textbf{MSE($\hat{\sigma_\epsilon}^2$).VS} & \textbf{MSE($\hat{\sigma_\epsilon}^2$).GR} & \textbf{MSE($\hat{\rho}^2$).VS} & \textbf{MSE($\hat{\rho}^2$).GR} \\ 
  \hline
  \hline
 \multirow{2}{*}{5} & 500 & 6.03 & 9.64e+22 & 1.38 & 3.95e+22 \\ 
   & 1000 & 1.08 & 1.96e+23 & 0.11 & 1.77e+23 \\ \hline
  \multirow{2}{*}{50} & 500 & 48.24 & 1.22e+23 & 7.14 & 1.38e+23 \\ 
   & 1000 & 3.27 & 5.39e+22 & 0.68 & 1.75e+23 \\ \hline
  \multirow{2}{*}{100} & 500 & 43.69 & 2.44e+22 & 9.44 & 9.19e+22 \\ 
   & 1000 & 1.26 & 1.62e+21 & 0.18 & 2.40e+21 \\ 
   \hline
   \toprule
\end{tabular}%
}
\end{table} 
The results are surprisingly in favor of the VS-based methodology. Clearly, the robust REML technique has completely failed to provide reasonable covariance estimates. We can not provide the results for all other choices of $n$ and noise model parameters because, the REML estimation in georob has huge time complexity. For example, for $\sigma_A = 5$ and $n = 1000$, on an average georob takes $145$ sec., whereas, VS-based estiamtion takes $2.5$ sec. -- i.e. georob takes on an average $58$ times more. Hence, both in terms of computational complexity and the accuracy of estimation, VS-based technique outperforms robust-REML method (as implemented in \citealt{georob18}).

In the next section, we are going to apply VS-based estimation technique on the coal ash data (\citealt{coalash}) and compare our results with that of \cite{georobMan}, where the author has implemented robust REML estimation on the same data set.

\section{Example: Coalash data}
\label{sec:readl-dat-ex}
\begin{figure}
  \centering
 \begin{subfigure}{0.48\textwidth}
     \centering
    \includegraphics[trim={2.5cm 1cm 2.5cm 1.5cm}, width=.55\textwidth, height=0.15\textheight]{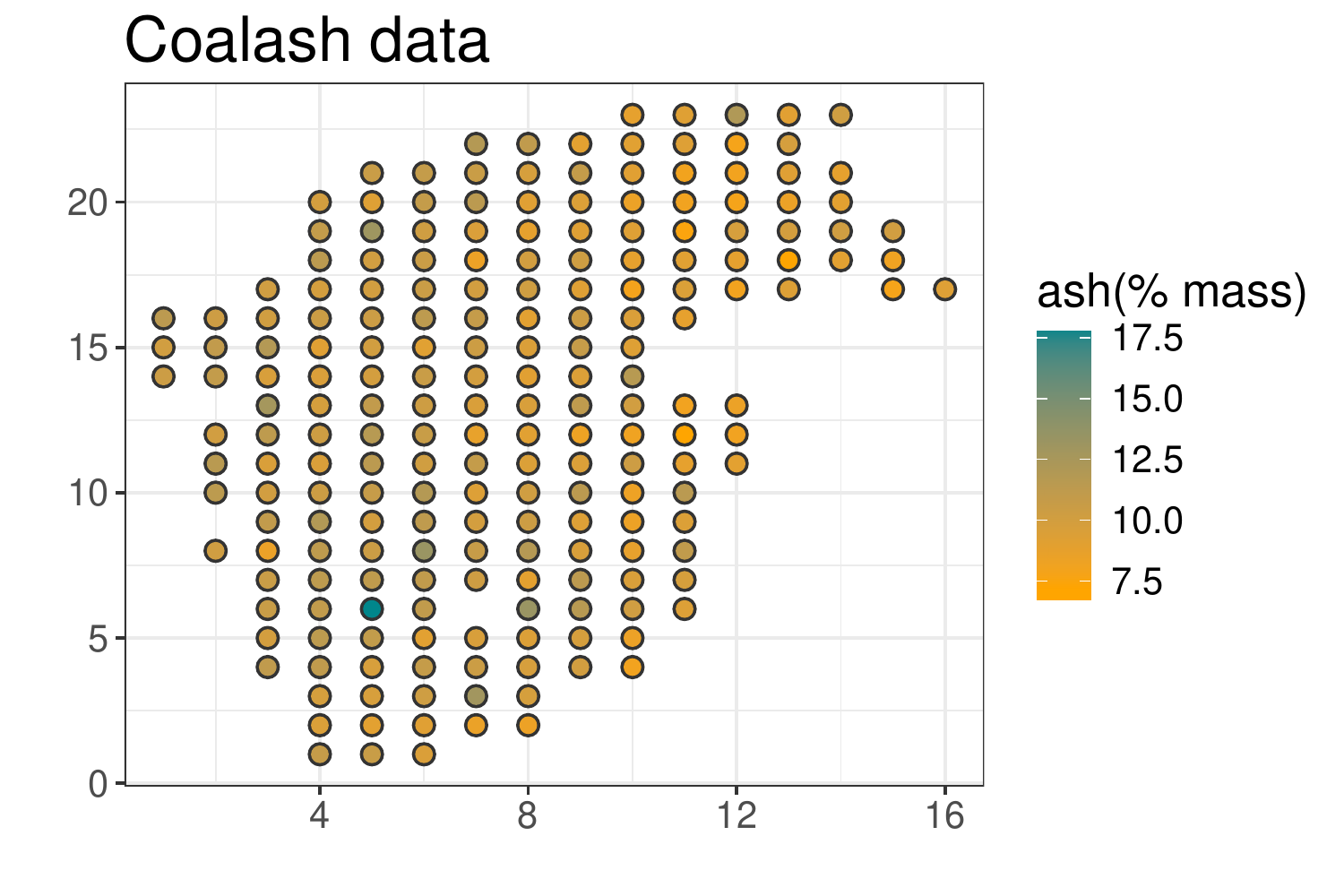}
    \subcaption{Ash \% in coal seams}\label{fig:CoalashDat}
 \end{subfigure}\hspace{2mm}%
 \begin{subfigure}{0.48\textwidth}
     \centering
    \includegraphics[trim={2.5cm 1cm 2.5cm 1.5cm}, width=.55\textwidth, height=0.15\textheight]{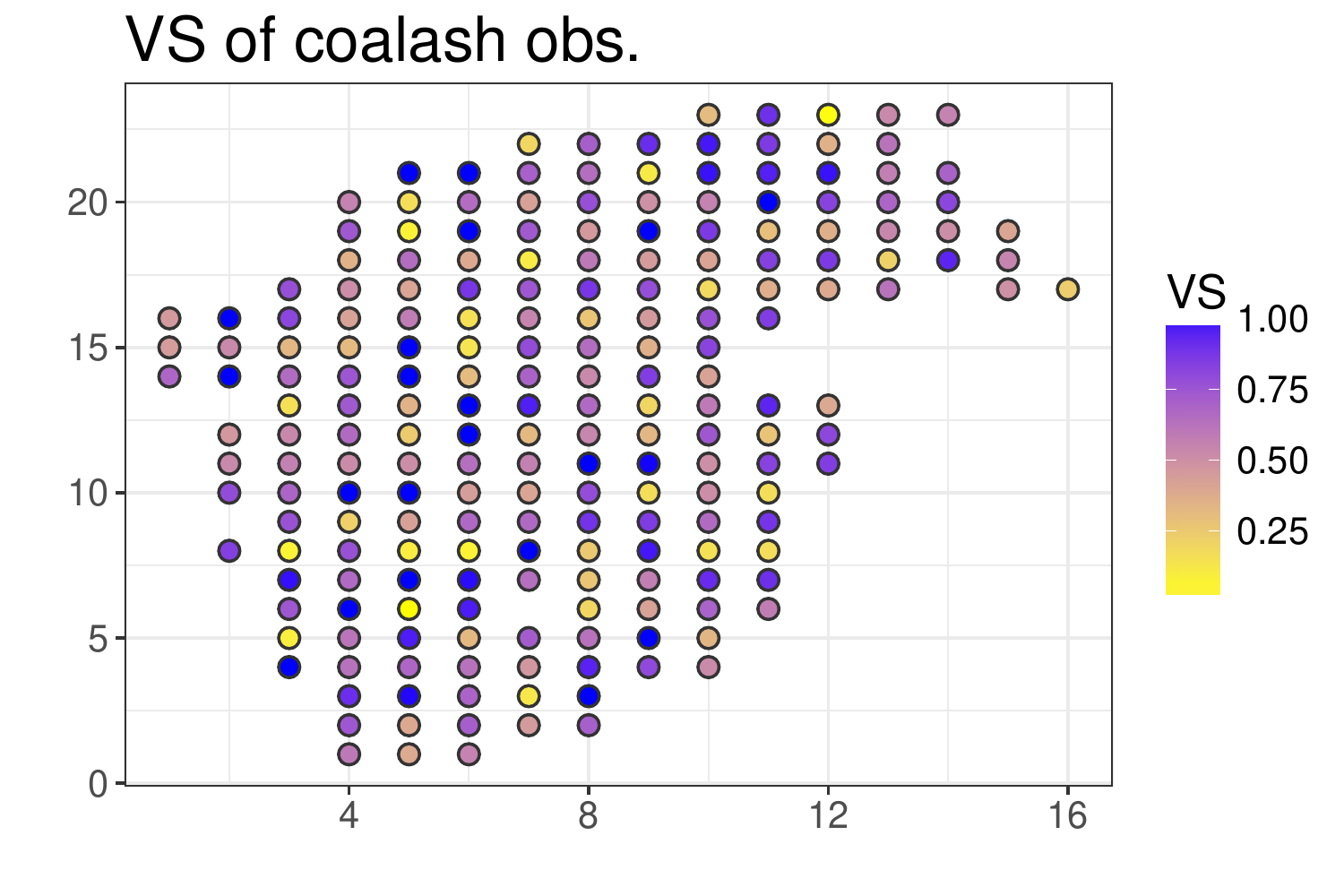}
    \subcaption{VS of the observations}\label{fig:vs-coalash}
 \end{subfigure}
 \caption{Spatial plots of the coalash data (a) and VS of the observations (b).}\label{fig:coalashDat-VS}
 \end{figure}

In this section we consider the coal ash data (\citealt{coalash}), which reports measurement (in $\%$ mass) of ash present in coal seam in Pennsylvania. The data set is available in R-package \texttt{gstat} (\citealt{gstat}) and has been previously used to demonstrate robust geostatistical techniques by \cite{cressie93} and \cite{georobMan}. Let us denote the target process as $\LP Y(\bfs) \RP$ where $Y(\bfs)$ is the mass percentage (often referred as \textit{$\%$ mass}) of ash in the coal seam at location $\bfs$. The observations plotted in Figure~\ref{fig:CoalashDat} are denoted by $\LP Z(\bfs_1), \dots , Z(\bfs_n) \RP$ where $n = 208$. Some of these observations are corrupted due to measurement errors or other types of unknown noises.

We first compute the VS (Median-VS) of the observations and plot it spatially in Figure~\ref{fig:vs-coalash}.
 \cite{cressie93} identified a set of observations as outliers by investigating the deviation of the observations from their overall sample median. VS uses the similar idea -- instead of using the overall median, it assigns a score to each of the observations according to its deviation from the `local' median relative to the `local' variation. The observations that were identified as outliers by \cite{cressie93} and \citealt{georobMan} receive VS less than 0.18 (in a scale of $(0,1]$). As both in theoretical analysis and simulation studies we have established that VS-based estimation is hampered if proportion of noisy observations is significantly large, we check the proportion of observations which have VS less than $0.22$ -- which implies the deviation from local median is at least $1.51 \times$local-IQR. We find that approximately $12\%$ of the observations have `high' noise associated with it.
 
Once the VS of the observations has been computed we incorporate those score to estimate the mean and covariance structure of the process robustly. Through some explanatory data analysis similar to \cite{georobMan}, we consider the following spatial regression model for the target process $\LP Y(\bfs) \RP$,
\begin{equation}
\label{eq:coalash-sp-reg}
    \begin{split}
        Y(\bfs) = \beta_0 + \beta_x s_x + \epsilon(\bfs),
    \end{split}
\end{equation} where $\bfbeta = \Lp \beta_0, \beta_x \Rp^\prime$ is the regression parameter, $s_x$ is longitude (i.e. $\bfs = \Lp s_x, s_y \Rp^\prime$) and $\LP \epsilon(\bfs) \RP$ is a zero mean second order stationary process with Mat\'ern covariance and the covariance parameter vector is given by $\bftheta = \Lp \sigma_\epsilon^2, \rho, \eta^2, \nu \Rp^\prime$ where $\sigma_\epsilon^2$ is the partiall sill, $\rho$ is the range, $\eta^2$ is the nugget and $\nu$ is the smoothness parameter (for details, see \citealt{haskard07}; \citealt{gelfand10}).

VS-based regression yields an estimate of $\hat{\bfbeta}_{\text{vs}} = \Lp 11.071, -0.188 \Rp^\prime$ for the regression parameter -- which suggests that the ash proportion decreases as we go from west to east.
\begin{figure}
  \centering
 \begin{subfigure}{0.48\textwidth}
     \centering
    \includegraphics[trim={3cm 1cm 1.5cm 0cm}, width=.5\textwidth, height=0.12\textheight]{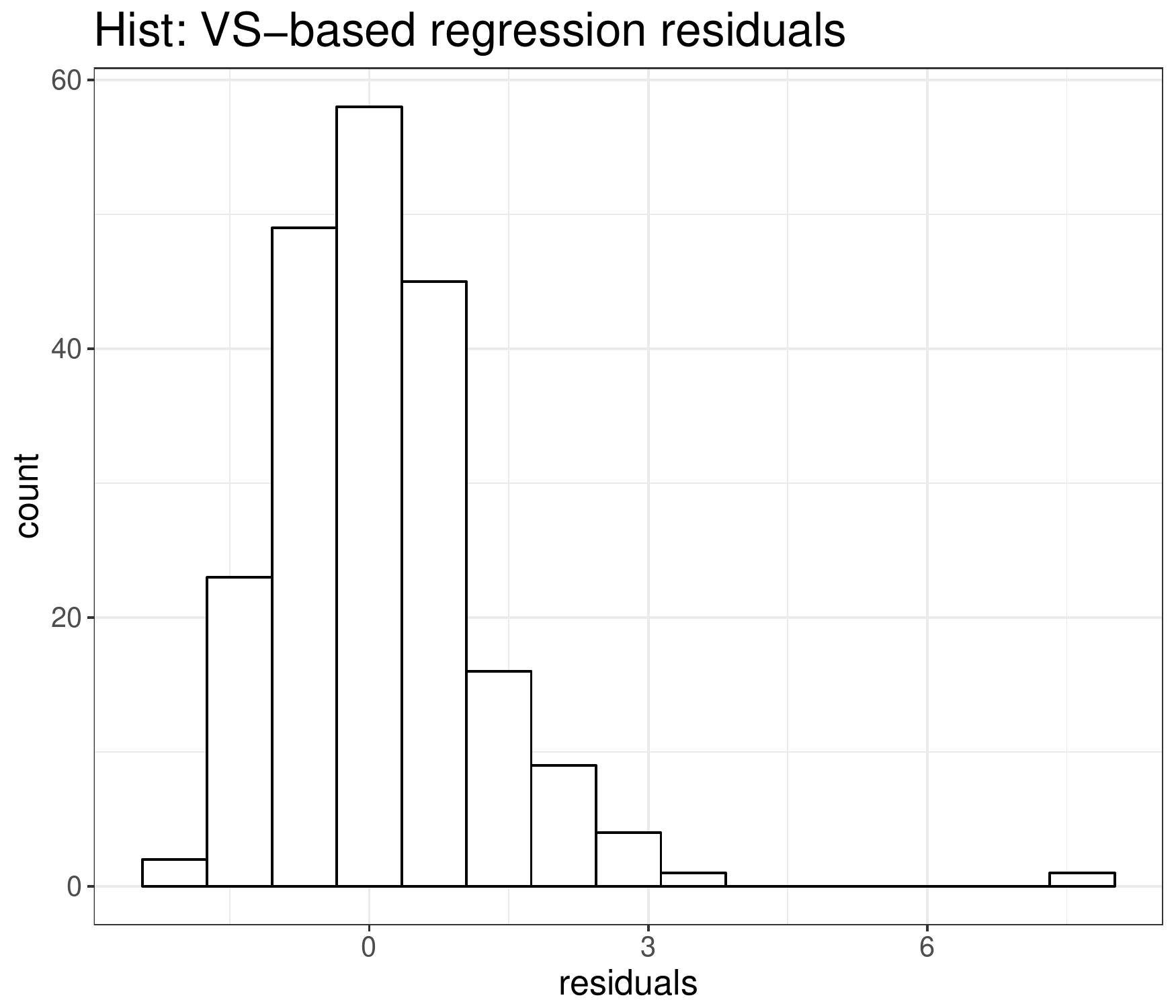}
    \subcaption{}\label{fig:hist-resid-vs}
 \end{subfigure}\hspace{2mm}%
 \begin{subfigure}{0.48\textwidth}
     \centering
    \includegraphics[trim={3cm 1cm 1.5cm 0cm}, width=.5\textwidth, height=0.12\textheight]{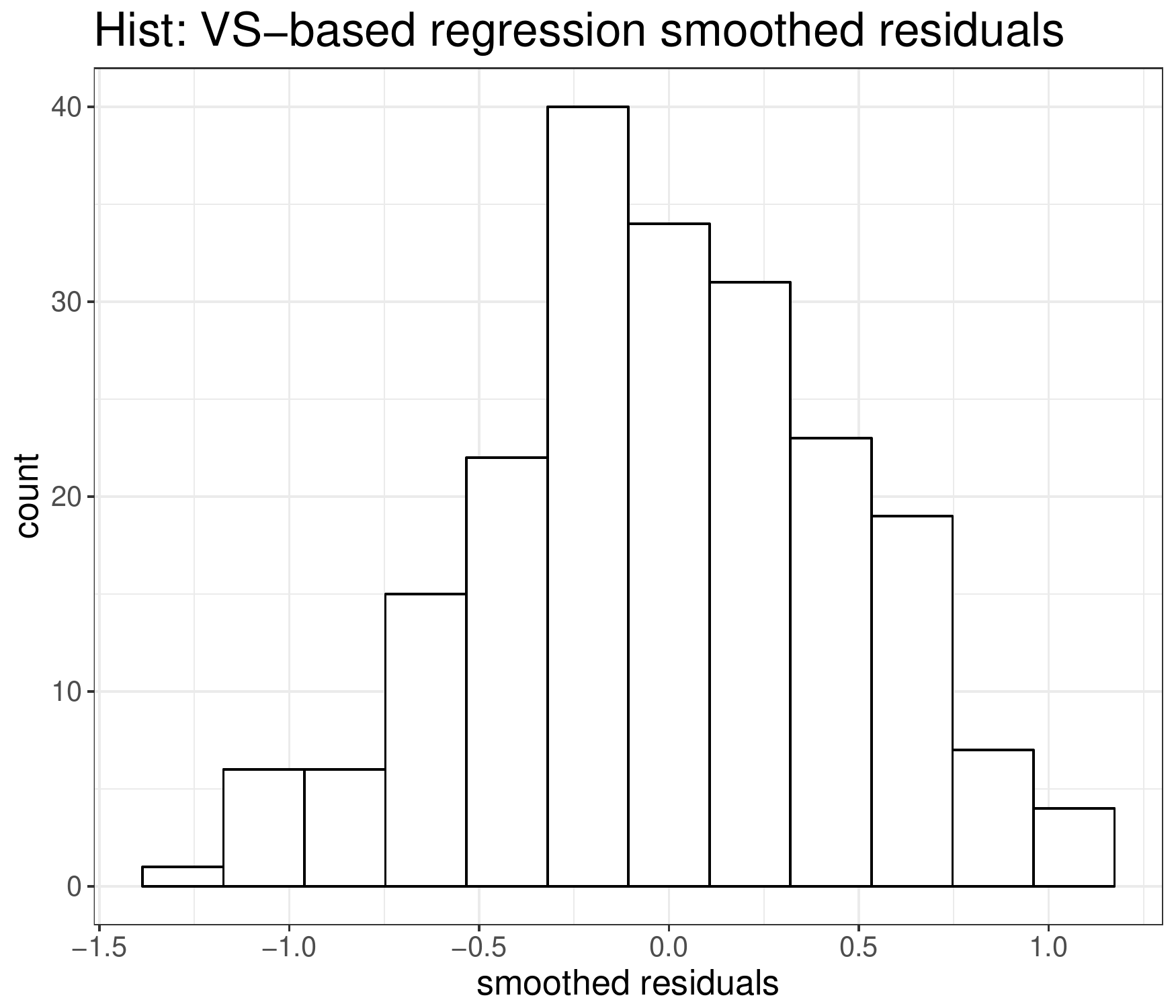}
    \subcaption{}\label{fig:hist-resid-vs-smooth}
 \end{subfigure}
\caption{VS-based smoothing of residuals: histogram of observed residuals from VS-based regression (a), histogram of smoothed residuals (b).}\label{fig:VS-Smoothing}
\end{figure} Next, the VS-based smoothing is implemented on the observed residuals from the VS-based regression and the effect of smoothing has been displayed through frequency plots in Figure~\ref{fig:VS-Smoothing}. Finally, the VS-based smooth residuals are used to estimate the covariance parameters through VS-based variogram model fitting. We summarize the details of VS-based covariance estimation in Figure~\ref{fig:coalashCovEst} and Table~\ref{tab:coalashCovEst}. Observe that, the range parameter estiamte ($0.486$) in Table~\ref{tab:coalashCovEst} is significantly greater than $0$, i.e. the VS-based analysis can capture the spatial correlation in the residual process, as opposed to the pure nugget model found in the analysis of \cite{georobMan}.\\
\begin{minipage}[c]{0.55\textwidth}
\centering
\resizebox{.6\columnwidth}{.18\textwidth}{%
\begin{tabular}{cc}
\hline
\hline
\textbf{Parameters} & \textbf{Estiamtes} \\ 
\hline
 partial sill ($\sigma_\epsilon^2$)    & 0.219              \\ 
range ($\rho$)        & 0.486              \\ 
nugget ($\eta^2$)     & 0.021               \\ 
smoothness ($\kappa$) & 5.00                  \\ 
   \hline
   \hline
\end{tabular}%
}
\captionof{table}{\small{Estimated Mat\'ern parameters.}}
\label{tab:coalashCovEst}
\end{minipage}
\begin{minipage}[c]{0.45\textwidth}
\includegraphics[trim={1.5cm 1.5cm 1cm 0.2cm}, width=.95\textwidth, height = .19\textheight]{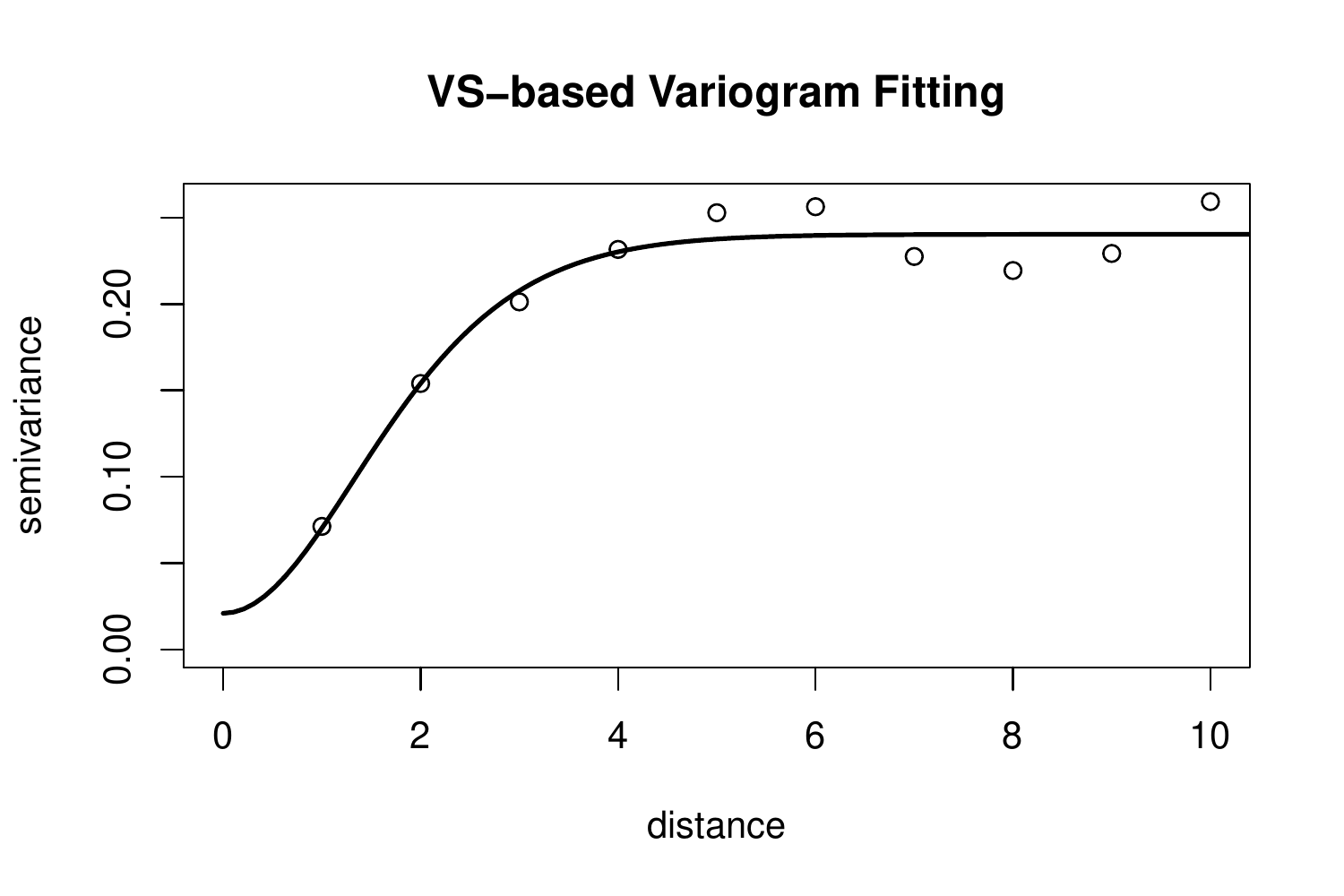}
\captionof{figure}{\small{Variogram estimation}}
\label{fig:coalashCovEst}
\end{minipage}

Next, we compare the VS-based analysis with the robust REML approach. To implement the robust REML methodology on the coal ash data the R-package \texttt{georob} (\citealt{georob18}) is used. We use leave-one-out cross-validation technique to compare the two approaches: for each of the observations in the coal ash data, we consider it as the test data and try to predict (kriging) it using all other observations without the test one. As, trying to predict a `bad' observation may lead to erroneous inference, from the test data set we exclude the observations marked as outliers by \cite{georobMan} (count $11$) in their analysis. But, the training data contains all the data points. Instead of using VS, we use the analysis done by \cite{georobMan} to remove the outliers from the test data so that the choice of test data in the cross-validated analysis is not biased towards the VS-based approach.

\begin{figure}
  \centering
  \begin{subfigure}{0.48\textwidth}
     \centering
    \includegraphics[trim={1.5cm 1.5cm 2.5cm .5cm}, width=.55\textwidth, height=0.14\textheight]{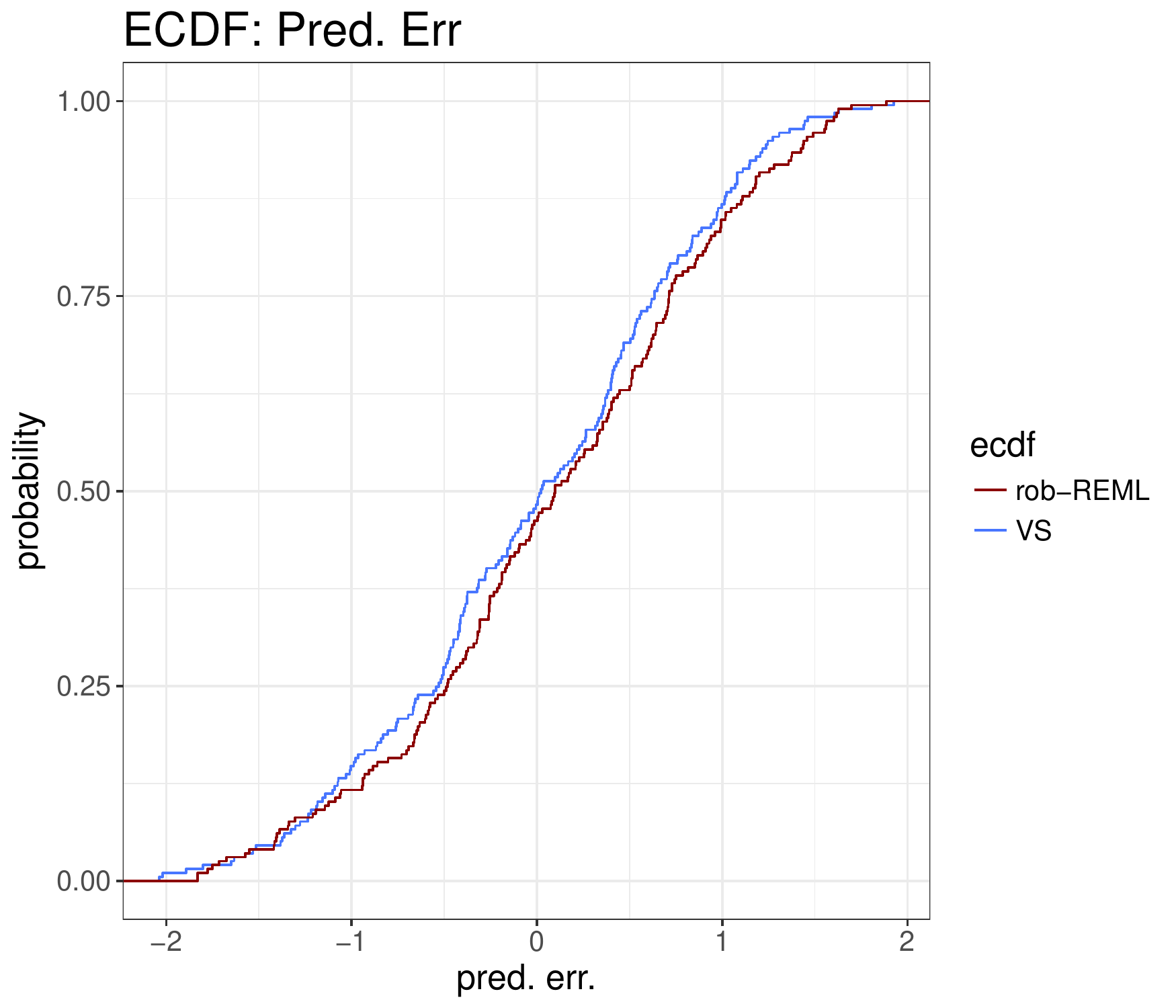}
    \subcaption{}\label{fig:ecdf-vs-gr}
 \end{subfigure}\hspace{2mm}%
 \begin{subfigure}{0.48\textwidth}
     \centering
    \includegraphics[trim={1.5cm 1.5cm 2.5cm .5cm}, width=.65\textwidth, height=0.14\textheight]{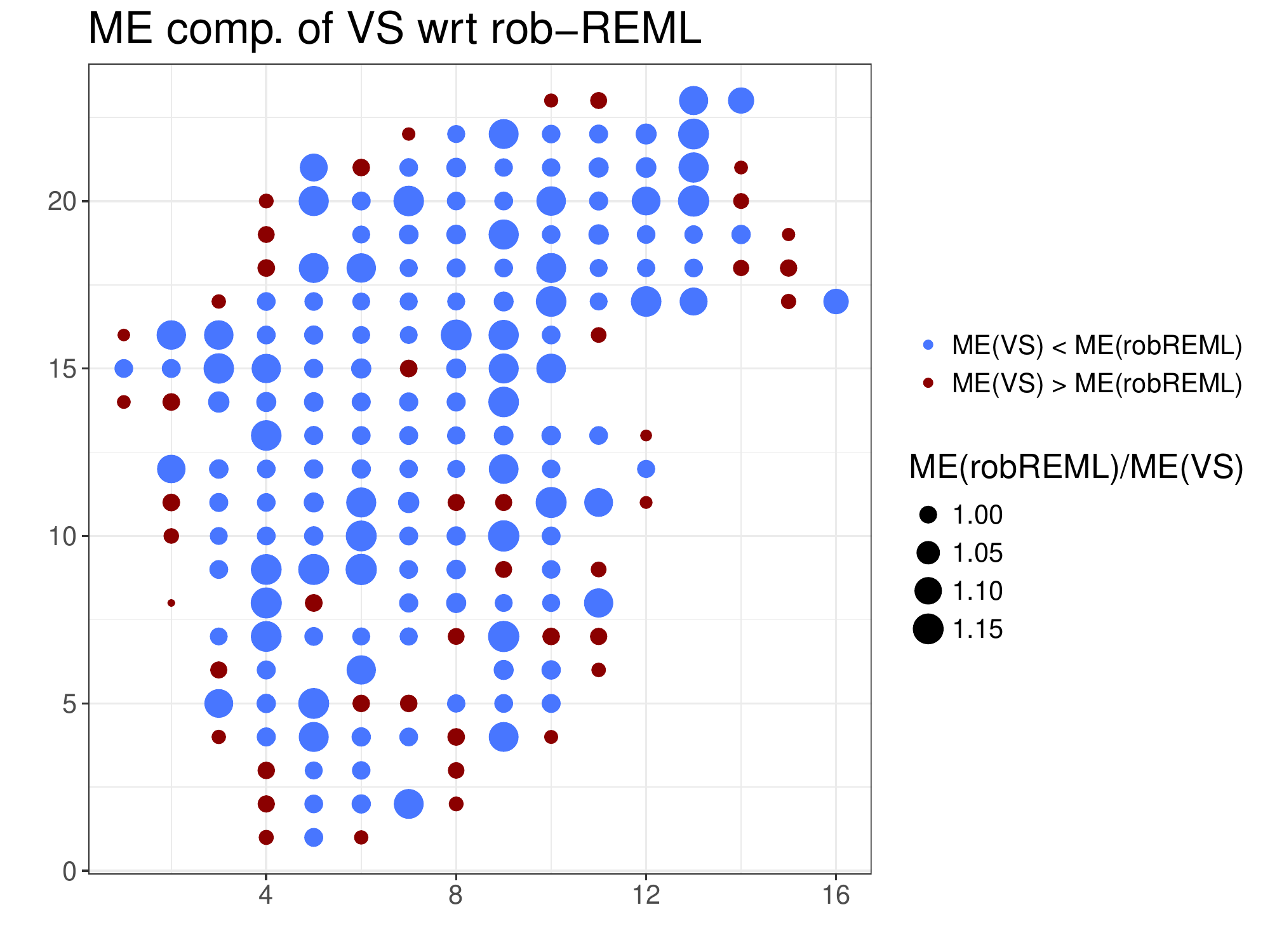}
    \subcaption{}\label{fig:re-vs-gr}
 \end{subfigure}
\caption{Prediction comparison between VS and robust-REML: (a) - empirical c.d.f. of prediction errors and, (d) - relative efficiency in terms of margin of prediction errors of VS w.r.t. robust-REML.}\label{fig:coalashPredDiag}
\end{figure}

In Figure~\ref{fig:coalashPredDiag}, we have summarized the results of comparative analysis between VS-based and robust-REML approach.
From the empirical distribution function plot in Figure~\ref{fig:ecdf-vs-gr} it is evident that the prediction errors corresponding to the robust REML predictor has higher mass in the tails than the VS-based predictions. Moreover, as we can see in Figure~\ref{fig:re-vs-gr}, the margins of prediction errors (ME), that is the half of the lengths of the prediction intervals, for the VS-based approach is either similar or smaller as compared to the robust REML for most of the test data points. To check the credibility of the VS-based prediction intervals we find that $98.5\%$ of the test data points were inside the $95\%$ VS-based prediction intervals. In terms of cross-validated mean squared prediction error (MSPE), the VS-based MSPE is $0.716$ whereas the robust-REML prediction has an MSPE of $0.743$, i.e. there is nearly $4\%$ gain in efficiency when VS-based analysis is carrired out in place of the robust REML approach. Moreover, in terms of the computational time VS-based approach has a huge advantage as compared to the robust REML, as shown in Table~\ref{tab:coalashCompTime}.
\begin{table}[ht]
\centering
\resizebox{0.35\columnwidth}{.1\textwidth}{%
\begin{tabular}{|c|cc|}
  \hline
Computation Time & VS-based & robust-REML \\ 
  \hline
Min. & 0.45 & 1.47 \\ 
  1st Qu. & 0.54 & 1.99 \\ 
  Median & 0.60 & 2.18 \\ 
  Mean & 0.61 & 8.24 \\ 
  3rd Qu. & 0.67 & 14.22 \\ 
  Max. & 0.93 & 29.75 \\ 
   \hline
\end{tabular}%
}
\caption{\small Time comparison between VS-based and robust-REML. Machine configuration: DELL R7425 Dual Processor AMD Epyc 32 core 2.2 GHz machines with 512GB RAM each running 64Bit Ubuntu Linux Version 18.04.}
\label{tab:coalashCompTime}
\end{table} Hence, for the analysis of larger spatial data sets, VS-based analysis has advantage over the robust REML approach.

\section{Conclusion}
\label{sec:conclusion}
In this paper, we have investigated the large sample behavior of the veracity scoring technique proposed by \cite{chak} for irregularly spaced spatial data when there is no reference data available. Through the asymptotic approximation of the VS we have established that VS of an observation is expected to take `small' value (close to $0$) if the noise variance associated with the observation is high and, it is going to take larger values if the associated noise variance is less. Moreover, under the mixed-increasing domain spatial asymptotic framework and for a non-stationary noise model, we have sheded light upon the circumstances under which the weighted least squares regression estimator with veracity scores as the corresponding weights -- which are data-driven and hence, spatially correlated -- is asymptotically consistent. We have also provided an order on the rate of convergence in probability for the VS-based regression estimator. Next, we have established that the asymptotic mean squared error of the VS-based estimation is highly resistant to the magnitudes of the noise variances associated with the corrupted observations. But, the accuracy of the estimation can be hampered if the proportion of noisy observations increases. In addition, we have also considered the asymptotic MSE of the ordinary least squares estimator and showed how it depends on both the proportion of noisy observations as well as the magnitudes of the noise variances. We have provided empirical justifications of the advantages of the VS-based analysis through rigorous simulations. Finally, we have implemented the VS-based approach on the coal ash data and compared the results with that of robust REML technique -- which shows that the VS-based approach has advantage both in terms of prediction accuracy and computational time.

There are several future directions to this work. First, the theoretical properties of the VS-based covariance estimation and kriging under similar spatial asymptotic framework can be explored. Second, a version of VS for spatio-temporal data will be of high interest, especially for analyzing real-time noisy crowdsourced data. Third, the results of this paper can be extended for spatial processes on $\rd$ with $d \geq 3$. That way the proerties of the VS and the VS-based estimation will be extended for spatio-temporal data as well.

\section*{Acknowledgements}
This research is partially funded by National Science Foundation (NSF) grant DMS-1613192. The authors would like to thank Dr. Alyson Wilson for bringing the importance of this work to authors' attention and also for her insightful remarks during the course of this work.

\appendix
\section{Proofs}
\label{sec:proofs}
To prove the main results of this paper we need to state couple of lemmas in the next subsection.
\subsection{Auxiliary lemmas}
\label{app-subsec:aux-lem}
Let us define $\hat{G}_{n(i)}(x) = 1 - \hat{F}_{n(i)}(x)$ and $\bar{G}_{n(i)}(x) = 1 - \bar{F}_{n(i)}(x)$.
\begin{lem}
\label{lemma:strong-mixing}
The strong-mixing coefficient $\alpha_w\Lp \cdot , \cdot \Rp$ of $\LP w(\bfs) \RP$ is bounded above by the strong-mixing coefficient $\alpha_\epsilon\Lp \cdot , \cdot \Rp$ of $\LP \epsilon(\bfs) \RP$, i.e.
$$
\alpha_w \Lp u, v \Rp \leq \alpha_\epsilon \Lp u , v \Rp,
$$ for any $u > 0$ , $v >0$.
\end{lem}

\begin{lem}
\label{lem:quantile-consist}
For any $0 < p < 1$,
\begin{equation}
\label{eq:quantile-consist}
\begin{split}
M_{n(i)}^p = \xi_{n(i)}^p + \frac{p - \hat{F}_{n(i)}(\xi_{n(i)}^p)}{\bar{f}_{n(i)}(\xi_{n(i)}^p)} + R_{n(i)},
\end{split}   
\end{equation} where $\bar{f}_{n(i)}(x) = \Lp \sum_{j = 1}^{n(i)} f_{i_j}(x) \Rp/n(i)$ and $\sqrt{n(i)}R_{n(i)} \to 0$ in probability. 
\end{lem}
Clearly, the representation in Equation~\ref{eq:quantile-consist} is the Ghosh-Bahadur representation in our setting. We extended the proofs of \cite{sen68} and \cite{jkg71} for heteroscedastic $\alpha$-mixing sequence of random variables. The proofs of these two lemmas have been provided in Section A.1 of the supplementary material.

\subsection{Proofs of the main results}
We start by stating the proof of Proposition~\ref{prop:MedVSApprox}.
\subsubsection{Proof of Proposition~\ref{prop:MedVSApprox}}
\proof Note that, 
\begin{equation}
\label{eq:decomp of median}
\begin{split}
Q_2\Lp \mathbf{Z}_i \Rp &= Q_2\Lp \Lp \bfx(\bfs_{i_1})^\prime \bfbeta + w(\bfs_{i_1}), \dots , \bfx(\bfs_{i_{n(i)}})^\prime \bfbeta + w(\bfs_{i_{n(i)}})  \Rp^\prime \Rp\\
&= \bfx(\bfs_{i})^\prime \bfbeta + Q_2\Lp \underbrace{\Lp w(\bfs_{i_1}), \dots , w(\bfs_{i_{n(i)}}) \Rp^\prime}_{ = \mathbf{w}_i} \Rp + O(\delta_n \lambda_n^{-1}),
\end{split}
\end{equation} 
as, from condition (C.3), 
\begin{equation}
\label{eq:decomp of covaraite}
\begin{split}
   \Lp \bfx(\bfs_i) - \bfx(\bfs_{i_j}) \Rp^\prime \bfbeta &= \sum_{k = 1}^p  \beta_k\Lp x_k(\bfs_i) - x_k(\bfs_{i_j}) \Rp \\ &= \sum_{k = 1}^p \beta_k \Lp \nabla x_k(\bfs^*_{i_j}) \Rp^\prime \Lp \bfs_i - \bfs_{i_j} \Rp = O\Lp \delta_n \lambda_n^{-1} \Rp,
\end{split}
\end{equation}
for some $\bfs^*_{i_j} \in \rtwo$ in between $\bfs_i$ and $\bfs_{i_j}$ component wise. Now, from Lemma~\ref{lem:quantile-consist} and condition (C.9), we have $Q_2(\mathbf{w}_i) = O_p\Lp(n(i))^{-1/2}\Rp = O_p(n^{-1/2} \lambda_n \delta_n^{-1})$. So, from equation~\ref{eq:decomp of median}, we have $Z(\bfs_i) - Q_2(\mathbf{Z}_i) = w(\bfs_i) + O_p(n^{-1/2} \lambda_n \delta_n^{-1} + \delta_n \lambda_n^{-1})$. Similar argument using the first and third sample quartile we can show that $\text{IQR}(\mathbf{Z}_i) = \mathcal{I}_n(\bfs_i) + O_p(n^{-1/2} \lambda_n \delta_n^{-1} + \delta_n \lambda_n^{-1})$. Now, through simple probability calculations we have
$$
\begin{aligned}
V^{(m)}(\bfs_i) &= \exp \Lp - \frac{\lvert Z(\bfs_i) - Q_2(\mathbf{Z}_i) \rvert}{\text{IQR}(\mathbf{Z}_i)} \Rp \\
&= \exp \Lp - \frac{\lvert w(\bfs_i) + O_p(a_n) \rvert}{\mathcal{I}_n(\bfs_i) + O_p(a_n)} \Rp \\
&= \exp \Lp - \frac{\lvert w(\bfs_i)\rvert}{\mathcal{I}_n(\bfs_i)} \Rp  + O_p(a_n),
\end{aligned}
$$ and hence proved.
\endproof

\subsubsection{Proof of Proposition~\ref{prop:IqrBound}}
\proof By definition of $\xi_{n(i)}^{p}$,
$
\bar{F}_{n(i)}(\xi_{n(i)}^{p}) = p
$. Now, observe that, $\frac{1}{n(i)}\sum_{j = 1}^{n(i)} F_{i_j}(\xi_{n(i)}^{p}) = \frac{1}{n(i)}\LP \sum_{j \in G_{i,n}} F_{i_j}(\xi_{n(i)}^{p}) + \sum_{j \in G_{i,n}^c} F_{i_j}(\xi_{n(i)}^{p})  \RP$, where $G_{i,n} \subset G_n$ is the set of indices of the observations in the square neighborhood around $\bfs_i$ which are not affected by the additive-multiplicative noise. From condition (C.1) and (C.2), $\lvert G_{i,n} \rvert/n(i) \geq q_e$. Hence,
$$
\begin{aligned}
p \geq \frac{1}{n(i)} \sum_{j \in G_{i,n}} F_{i_j}(\xi_{n(i)}^{p}) \geq q_e F_\epsilon(\xi_{n(i)}^{p}/ \sigma_\epsilon),
\end{aligned}
$$ which proves that $\xi_{n(i)}^{p} \leq \sigma_\epsilon  \Lp F_\epsilon^{-1}(\text{min}\LP p/q_e, 1 \RP)\Rp$. In place of $p$, using the similar argument for the $(1-p)$-th quantile we can show that $\xi_{n(i)}^{p} \geq \sigma_\epsilon  \Lp F_\epsilon^{-1}(\text{max}\LP 1 - \frac{1-p}{q_e}, 0 \RP)\Rp$. Taking $p = 0.75$ and $0.25$ we have the proof.
\endproof

\subsubsection{Proof of Theorem~\ref{theo:VS-Mean-Prepresentation}}
\label{app-subsubsec:med-vs-reg-rep}
\proof From definition of Median-VS-based regression estimator,
$$
\hat{\bfbeta}^{(m)}_{\text{vs}} - \bfbeta \; = \; \Lp \frac{1}{n} \bfX^\prime \mathbf{D}_v^{(m)} \bfX \Rp^{-1} \Lp \frac{1}{n} \bfX^\prime \mathbf{D}_v^{(m)} \mathbf{w} \Rp,
$$ where $\mathbf{D}_v^{(m)} = \text{diag}\Lp V^{(m)}(\bfs_1), \dots , V^{(m)}(\bfs_n) \Rp$. For simplicity of notation, as we are only considering Median-VS here, let us denote $\mathbf{D}_v^{(m)}$ as $\mathbf{D}_v$. Now, from Proposition~\ref{prop:MedVSApprox}, $
\frac{1}{n} \bfX^\prime \mathbf{D}_v \mathbf{w} = \frac{1}{n} \bfX^\prime \tilde{\mathbf{D}}_v \mathbf{w} + O_p(a_n)
$ and $
\frac{1}{n} \bfX^\prime \mathbf{D}_v \bfX = \frac{1}{n} \bfX^\prime E\Lp \tilde{\mathbf{D}}_v\Rp \bfX + \frac{1}{n} \bfX^\prime \Lp \tilde{\mathbf{D}}_v - E\Lp \tilde{\mathbf{D}}_v \Rp \Rp \bfX + O_p(a_n).
$ Next we will show that $\frac{1}{n} \bfX^\prime \Lp \tilde{\mathbf{D}}_v - E\Lp \tilde{\mathbf{D}}_v \Rp \Rp \bfX = O_p(\lambda_n^{-1})$.

\hspace{6mm} Notice that $\frac{1}{n} \bfX^\prime \Lp \tilde{\mathbf{D}}_v - E\Lp \tilde{\mathbf{D}}_v \Rp \Rp \bfX$ is a $p\times p$ matrix and hence enough to show that a general element of of the matrix say $(k,l)$-th element is $O_p(\lambda_n^{-1})$. Now, from simple calculations,  
$$
\begin{aligned}
\Lp\frac{1}{n} \bfX^\prime \Lp \tilde{\mathbf{D}}_v - E\Lp \tilde{\mathbf{D}}_v \Rp \Rp \bfX\Rp_{kl} &= \frac{1}{n} \sum_{i=1}^n x_k\Lp \bfs_i \Rp x_l\Lp \bfs_i \Rp \LP \tilde{V}^{(m)}(\bfs_i) - E \Lp \tilde{V}^{(m)}(\bfs_i) \Rp \RP\\
&= \frac{1}{n}\sum_i T_{kl}(\bfs_i),
\end{aligned}
$$ where, $T_{kl}(\bfs_i) = x_k\Lp \bfs_i \Rp x_l\Lp \bfs_i \Rp \LP \tilde{V}^{(m)}(\bfs_i) - E \Lp \tilde{V}^{(m)}(\bfs_i) \Rp \RP$. As, $E\Lp T_{kl}(\bfs_i) \Rp = 0$ enough to show that $\var\Lp \frac{1}{n}\sum_i T_{kl}(\bfs_i) \Rp = O(\lambda_n^{-2})$. To do so we will adopt the similar strategy used in the proof of Lemma~\ref{lem:quantile-consist}. We will divide the whole sampling region $\mathcal{D}_n = \lambda_n[0,1]^2$ into disjoint unit squares $\LP B_1, \dots , B_{b_n} \RP$ and $b_n$ is the minimal natural number such that $\mathcal{D}_n \subset \cup_{m = 1}^{b_n} B_m$. Clearly $b_n = \lceil \lambda_n \rceil^2$ and we can rewrite the set of unit squares as $\LP B_m : \; m \in \LP 1, \dots , b_n \RP \RP = \LP B_{m_1, m_2} : \; m_1, m_2 \in \LP 1, \dots , \sqrt{b_n}= \lceil \lambda_n \rceil \RP \RP$, i.e. by $m_1$ and $m_2$ we are indexing the squares row and column wise respectively. Then following similar arguments we can show that $E\Lp \frac{1}{n}\sum_i T_{kl}(\bfs_i) \Rp^2 = O(\lambda_n^{-2})$. Hence, $
\frac{1}{n} \bfX^\prime \mathbf{D}_v \bfX = \frac{1}{n} \bfX^\prime E\Lp \tilde{\mathbf{D}}_v\Rp \bfX + O_p(\lambda_n^{-1} + a_n) = \frac{1}{n} \bfX^\prime E\Lp \tilde{\mathbf{D}}_v\Rp \bfX + O_p(a_n)$. So,
$$
\begin{aligned}
\hat{\bfbeta}^{(m)}_{\text{vs}} - \bfbeta &=  \Lp \frac{1}{n} \bfX^\prime \mathbf{D}_v \bfX \Rp^{-1} \Lp \frac{1}{n} \bfX^\prime \mathbf{D}_v \mathbf{w} \Rp\\
&= \Lp \frac{1}{n} \bfX^\prime E\Lp \tilde{\mathbf{D}}_v\Rp \bfX + O_p(a_n) \Rp^{-1} \Lp \frac{1}{n} \bfX^\prime \tilde{\mathbf{D}}_v \mathbf{w} + O_p(a_n) \Rp\\
&= \Lp \frac{1}{n} \bfX^\prime E\Lp \tilde{\mathbf{D}}_v\Rp \bfX \Rp^{-1} \Lp \frac{1}{n} \bfX^\prime \tilde{\mathbf{D}}_v \mathbf{w}  \Rp + O_p(a_n),
\end{aligned}
$$and hence proved.
\endproof

\subsubsection{Proof of Corollary~\ref{cor:MedVSConsistent}}
\label{app-subsubsec:med-vs-consist-cor}
\proof From condition (C.11), $\frac{1}{n} \mathbf{X}^\prime \tilde{\mathbf{D}}_v \mathbf{w} = O_p\Lp \sqrt{\var\Lp \frac{1}{n} \mathbf{X}^\prime \tilde{\mathbf{D}}_v \mathbf{w} \Rp} \Rp$. Using similar argument as in the proof of Theorem~\ref{theo:VS-Mean-Prepresentation}, we can show that $\var\Lp \frac{1}{n} \mathbf{X}^\prime \tilde{\mathbf{D}}_v \mathbf{w} \Rp = O(\lambda_n^{-2})$. Hence, $\hat{\bfbeta}^{(a)}_{\text{VS}} = \bfbeta + \Lp \frac{1}{n} \mathbf{X}^\prime E\Lp \tilde{\mathbf{D}}_v \Rp \mathbf{X} \Rp^{-1} O_p \Lp \lambda_n^{-1} \Rp + O_p(a_n)$. Now, for all $i \in \LP 1, \dots , n \RP$,
$$
\begin{aligned}
\Lp E\Lp \tilde{\mathbf{D}}_v \Rp \Rp_{ii} &= E\Lp \exp\Lp -\frac{\lvert w(\bfs_i) \rvert}{\mathcal{I}_n(\bfs_i)} \Rp \Rp\\
&= \int_0^\infty P\Lp\exp\Lp -\frac{\lvert w(\bfs_i) \rvert}{\mathcal{I}_n(\bfs_i)} \Rp > x\Rp dx\\
&= \int_0^\infty e^{-y}P\Lp \lvert w(\bfs_i) \rvert <  \mathcal{I}_n(\bfs_i) y\Rp dy\;\;\;\; [\text{taking}\; x = e^{-y}]\\
&> \int_0^\infty e^{-y} P(\vert \epsilon(\bfs_i) \vert < y C_\epsilon^{(l)}(q_e))\; = \psi_\epsilon\Lp C_\epsilon^{(l)}(q_e) \Rp.\\
\end{aligned}
$$ Clearly, as from condition (C.12) $q_e > 0.75$, $C_\epsilon^{(l)}(q_e) > 0$ and under assumption (C.13) $\psi_\epsilon\Lp C_\epsilon^{(l)}(q_e) \Rp$ is strictly positive. Note that, $\frac{1}{n}\mathbf{X}^\prime \mathbf{X} \succeq \frac{1}{n} \mathbf{X}^\prime E\Lp \tilde{\mathbf{D}}_v \Rp \mathbf{X} \succ \frac{\psi_\epsilon\Lp C_\epsilon^{(l)}(q_e) \Rp}{n}\mathbf{X}^\prime \mathbf{X}$ where, from condition (C.10), $\frac{1}{n}\mathbf{X}^\prime \mathbf{X} \to \bfC_X$ for some positive definite matrix $\bfC_X$. Clearly, for large enough $n$, $\Lp\frac{1}{n} \mathbf{X}^\prime E\Lp \tilde{\mathbf{D}}_v \Rp \mathbf{X}\Rp^{-1} \preceq \psi_\epsilon\Lp C_\epsilon^{(l)}(q_e) \Rp \bfC_X$ and hence we have the proof. \endproof

\subsubsection{Proof of Theorem~\ref{theo:vs_ineq}}
\label{app-subsubsec:vs-ineq}
\proof Observe that,
$$
\begin{aligned}
E\Lp \lVert \mathbf{l}_n^{\text{vs}} \rVert^2 \Rp &= E\Lp \mathbf{w}^\prime \mathbf{\tilde{D}}_v  \bfX \Lp \bfX^\prime E\Lp \tilde{\mathbf{D}}_v \Rp \bfX \Rp^{-2} \bfX^\prime  \mathbf{\tilde{D}}_v \bfw \Rp\\
&= \frac{1}{n}E\Lp \bfw^\prime \tilde{\mathbf{D}}_v \bfA_n \tilde{\mathbf{D}}_v \bfw \Rp \quad \text{where,} \;\;\frac{1}{n} \bfX^\prime E\Lp \tilde{\mathbf{D}}_v \Rp \bfX = \mathbf{B}_n, \frac{1}{n}\bfX \mathbf{B}_n^{-2} \bfX^\prime = \bfA_n \\
&= \frac{1}{n}\text{tr}\Lp \bfA_n E\Lp \tilde{\mathbf{D}}_v \bfw \bfw^\prime \tilde{\mathbf{D}}_v \Rp \Rp\\
&= \frac{1}{n}\sum_{i = 1}^n (\bfA_n)_{ii} E\Lp \tilde{V}(\bfs_i)^2 w(\bfs_i)^2 \Rp + \frac{1}{n}\sum_{i \neq j} (\bfA_n)_{ij} E\Lp \tilde{V}(\bfs_i) w(\bfs_i) \tilde{V}(\bfs_j) w(\bfs_j) \Rp
\end{aligned}
$$ We can show that $\frac{1}{n}\bfX^\prime \bfX \succeq \mathbf{B}_n \succ \psi_\epsilon\Lp C_\epsilon^{(l)}(q_e) \Rp\frac{1}{n}\bfX^\prime \bfX$ where $\psi_\epsilon\Lp C_\epsilon^{(l)}(q_e) \Rp > 0$ and from condition (C.10) $\frac{1}{n}\bfX^\prime \bfX \to \bfC_X$ as $n \to \infty$. Hence, for large enough $n$,
$$
\lambda_{\text{min}}\Lp \mathbf{B}_n \Rp \geq \psi_\epsilon\Lp C_\epsilon^{(l)}(q_e) \Rp \lambda_{\text{min}}\Lp \bfC_X \Rp/2,
$$ and so,
$$
\Lp\bfA_n\Rp_{ii} \leq \frac{4 \;\text{sup}_i \lVert \bfX[i,] \rVert^2}{n \Lp \psi_\epsilon\Lp C_\epsilon^{(l)}(q_e) \Rp\lambda_{\text{min}}\Lp \bfC_X \Rp \Rp^2 };$$ and
$$
\begin{aligned}
E\Lp \tilde{V}(\bfs_i)^2 w(\bfs_i)^2 \Rp &= E\Lp \exp\Lp \frac{-2 \lvert w(\bfs_i) \rvert}{\mathcal{I}_n(\bfs_i)} \Rp w(\bfs_i)^2 \Rp\\
&= \sigma_i^2 E\Lp \exp \Lp -t_{i,n} \lvert u_i \rvert \Rp u_i^2 \Rp\\
&\leq \sigma_i^2 \exp(-2) 4/t_{i,n}^2 \;\; \text{where} \; t_{i,n} = 2\sigma_i/\mathcal{I}_n(\bfs_i)\\
&\leq \exp(-2)(C_\epsilon^{(u)}(q_e))^2.
\end{aligned}
$$ So, we have shown that,
$$
\frac{1}{n}\sum_{i = 1}^n (\bfA_n)_{ii} E\Lp \tilde{V}(\bfs_i)^2 w(\bfs_i)^2 \Rp = O\Lp n^{-1}\Lp C_\epsilon^{(u)}(q_e)\Rp^2\Lp\psi_\epsilon\Lp C_\epsilon^{(l)}(q_e) \Rp\Rp^{-2}\Rp.
$$ Using the similar trick as used in the proof of Lemma~\ref{lem:quantile-consist}, i.e. dividing the region into unit blocks and also using the fact that $\lvert \Lp \bfA_n \Rp_{ij} \rvert = O(n^{-1})$, we can show that
$$
\lvert \frac{1}{n}\sum_{i \neq j} (\bfA_n)_{ij} E\Lp \tilde{V}(\bfs_i) w(\bfs_i) \tilde{V}(\bfs_j) w(\bfs_j) \Rp \rvert = O(\lambda_n^{-4}),
$$ which concludes the proof.
\endproof

\subsubsection{Proof of Theorem~\ref{theo:ols_ineq}}
\label{app-subsubsec:ols-ineq}
\proof Recall, $\mathbf{l}_n^{\text{ols}} = \Lp n^{-1} \bfX^\prime \bfX \Rp^{-1} \Lp n^{-1} \bfX^\prime \mathbf{w} \Rp$. So,
$$
\begin{aligned}
E\Lp \lVert \mathbf{l}_n^{\text{ols}} \rVert^2 \Rp &= E\Lp \mathbf{w}^\prime   \bfX \Lp \bfX^\prime  \bfX \Rp^{-2} \bfX^\prime   \bfw \Rp \\
&= \frac{1}{n} \text{tr} \Lp \bfH_n \boldsymbol{\Sigma}_w \Rp\\
&=\frac{1}{n}\sum_{i = 1}^n (\bfH_n)_{ii} E\Lp w(\bfs_i)^2 \Rp + \frac{1}{n}\sum_{i \neq j} (\bfH_n)_{ij} E\Lp w(\bfs_i)  w(\bfs_j) \Rp,
\end{aligned}
$$ where $\bfH_n = n^{-1}\mathbf{X}(n^{-1}\mathbf{X}^\prime \mathbf{X})^{-2}\mathbf{X}^\prime$, $\boldsymbol{\Sigma}_w = \var\Lp \mathbf{w} \Rp$. Let us also denote $n^{-1}\bfX^\prime \bfX$ by $\bfC_n$ and by condition (C.10), $\bfC_n \to \bfC_X$ as $n \to \infty$. Note that $\text{tr}\Lp \bfH_n \Rp = \text{tr}\Lp \bfC_n^{-1} \Rp \to \text{tr}\Lp \bfC_X^{-1} \Rp = 2C^{(0)} > 0.$ So, for large enough $n$, $\text{tr}\Lp \bfH_n \Rp \geq C^{(0)} > 0$. Also, for large enough $n$,
$$
\Lp \bfH_n \Rp_{ii} \leq \frac{1}{n} \lambda_{\text{max}}\Lp \bfC_n^{-2} \Rp \lVert \bfX[i,] \rVert^2 < \frac{C^{(1)}}{n}.
$$ For any $\varepsilon > 0$ let us define,
$$\mathcal{M}_n(\varepsilon) = \LP 1 \leq i \leq n : \Lp \bfH_n \Rp_{ii} > n^{-1} \varepsilon \RP,$$ and let us denote the cardinality of this set by $m_n$ (dependent on $\varepsilon$). Now consider a $\varepsilon > 0$ such that $\varepsilon < \text{min}\Lp C^{(0)}, C^{(1)} \Rp$. Then,
$$
\begin{aligned}
C^{(0)} \leq \text{tr}\Lp \bfH_n \Rp = \sum_{i = 1}^{n} \Lp \bfH_n \Rp_{ii} \leq m_n \frac{C^{(1)}}{n} + (n - m_n) \frac{\varepsilon}{n},
\end{aligned}
$$ and by simple algebric calculation we have $m_n \geq \Lp C^{(0)} - \varepsilon \Rp \Lp C^{(1)} - \varepsilon \Rp^{-1}n$. Recall, $E(w(\bfs_i)^2)$ is equal to $\sigma_\epsilon^2$ if $i \in G_n$ and is equal to $ \sigma_\epsilon^2 + \tau_i^2$ otherwise. So,
$$
\begin{aligned}
\frac{1}{n}\sum_{i = 1}^n (\bfH_n)_{ii} E\Lp w(\bfs_i)^2 \Rp &= \frac{\sigma_\epsilon^2}{n}\sum_{i = 1}^n \Lp \bfH_n \Rp_{ii} + \frac{1}{n}\sum_{i \in G_n^c}  \tau_i^2 \Lp \bfH_n \Rp_{ii}\\
&> \frac{\sigma_\epsilon^2}{n}\sum_{i \in \mathcal{M}_n(\varepsilon)} \frac{\varepsilon}{n} + \frac{1}{n}\sum_{i \in \mathcal{M}_n(\varepsilon)\cap G_n^c} \tau_i^2 \frac{\varepsilon}{n}\\
& \geq  \frac{\varepsilon}{n}\Lp \frac{ \sigma_\epsilon^2 \Lp C^{(0)} - \varepsilon \Rp}{\Lp C^{(1)} - \varepsilon \Rp} + \frac{1}{n}\sum_{i \in \mathcal{M}_n(\varepsilon)\cap G_n^c} \tau_i^2 \Rp.
\end{aligned}
$$ Finally, using the similar methodology as in proof of Theorem~\ref{theo:vs_ineq} we can show that $\lvert\frac{1}{n}\sum_{i \neq j} (\bfH_n)_{ij} E\Lp w(\bfs_i)  w(\bfs_j) \Rp\rvert = O(\lambda_n^{-4})$. Hence we have shown that, for any $0 < \varepsilon < \text{min}\Lp C^{0}, C^{1} \Rp$ and large enough $n$,
$$E\Lp \lVert \mathbf{l}_n^{\text{ols}} \rVert^2 \Rp > n^{-1}\varepsilon\Lp \sigma_\epsilon^2 \Lp C^{(0)} - \varepsilon \Rp\Lp C^{(1)} - \varepsilon \Rp^{-1} + n^{-1}\sum_{i \in \mathcal{M}_n(\varepsilon)\cap G_n^c} \tau_i^2 \Rp - C_1 \lambda_n^{-4}.$$
\endproof

\subsubsection{Proof of Corollary~\ref{cor:ols_ineq}}
\label{app-subsubsec:ols-ineq-cor}
\proof Observe that, for large enough $n$, for all $i \in \LP 1, \dots , n \RP$, 
$$
\begin{aligned}
n\Lp\bfH_n\Rp_{ii} &= \Lp\bfX \bfC_n^{-2}\bfX^\prime\Rp_{ii}\\
&\geq \lambda_{\text{min}} \Lp \bfC_n^{-2} \Rp \lVert \bfX[i,] \rVert^2\\
& > \Lp\lambda_{\text{max}} \Lp \bfC_X\Rp\Rp^{-2} \underset{1\leq i \leq n}{\text{min}} \lVert \bfX[i,] \rVert^2 /2 \;\; \Lp= C_2, \; \text{say} \Rp.
\end{aligned}
$$ Clearly, from the assumption of the Corollary~\ref{cor:ols_ineq}, $C_2 > 0$. Take any $\varepsilon > 0$ such that $\varepsilon < C_2$. Then, for all $i \in \LP 1, \dots , n \RP$, $\Lp \bfH_n \Rp_{ii} > n^{-1}\varepsilon$ and hence, $$\frac{1}{n}\sum_{i = 1}^n (\bfH_n)_{ii} E\Lp w(\bfs_i)^2 \Rp > \frac{1}{n}\Lp \varepsilon\sigma_\epsilon^2  + \frac{1}{n}\sum_{i \in G_n^c} \tau_i^2 \Rp.$$ The rest of the proof is similar to that of Theorem~\ref{theo:ols_ineq}.
\endproof


\bibliographystyle{rss}
\bibliography{VS_JRSSB}
\end{document}